\documentclass[]{aa} 
\usepackage[varg]{txfonts}
\usepackage{latexsym} 
\usepackage{natbib}
\bibpunct{(}{)}{;}{a}{}{,} 
\usepackage{graphicx}  
\usepackage[colorlinks=true, citecolor=blue, plainpages=false]{hyperref}
\usepackage{amsbsy}
\usepackage{color,soul}
\usepackage{multirow}
\usepackage{footnote}
\usepackage{soul}

\begin{document}

\title{Correcting for chromatic stellar activity effects in transits with multiband photometric monitoring: Application to WASP-52}
\titlerunning{Correcting for chromatic stellar activity effects in transits}

\author{A.\,Rosich\inst{1,2}
       \and E.\,Herrero\inst{1,2}
       \and M.\,Mallonn\inst{3}
       \and I.\,Ribas\inst{1,2}
       \and J.\,C. Morales\inst{1,2}
       \and M.\,Perger\inst{1,2}
       \and G.\,Anglada-Escud\'e\inst{1,2}
       \and T.\,Granzer\inst{3}
       }
    
\offprints{Albert Rosich, \email{rosich@ice.cat}}

\institute{
\inst{1}Institut de Ci\`encies de l'Espai (ICE, CSIC), Campus UAB, Can Magrans s/n, 08193 Bellaterra, Spain\\
\inst{2}Institut d'Estudis Espacials de Catalunya (IEEC), 08034 Barcelona, Spain\\
\inst{3}Leibniz-Institut f\"ur Astrophysik Potsdam, An der Sternwarte 16, 14482 Potsdam, Germany\\
}

\date{Received January 27, 2020 / Accepted June 29, 2020}

\abstract
{The properties, distribution, and evolution of inhomogeneities on the surface of active stars, such as dark spots and bright faculae, significantly influence the determination of the parameters of an orbiting exoplanet. The chromatic effect they have on transmission spectroscopy, for example, could affect the analysis of data from future space missions such as James Webb Space Telescope (JWST) and Ariel.}
{To quantify and mitigate the effects of those surface phenomena, we developed a modelling approach to derive the surface distribution and properties of active regions by modelling simultaneous multi-wavelength time-series observables.}
{We present an upgraded version of the \textsf{StarSim} code, now featuring the capability to solve the inverse problem and derive the properties of the stars and their active regions by modelling time-series data. As a test case, we analyse $\sim$600\,days of $BVRI$ multiband photometry from the $0.8$-m Joan Oró (TJO) and $1.2$-m STELLA telescopes of the K2\,V exoplanet host star WASP-52. From the results, we further simulated the chromatic contribution of surface phenomena on the observables of its transiting planet.}
{Using \textsf{StarSim} we are able to determine the relevant activity parameters of WASP-52 and reconstruct the time-evolving longitudinal map of active regions. The star shows a heterogeneous surface composed of dark spots with a mean temperature of $575\pm150$\,K lower than the photospheric value, with filling factors ranging from 3 to 14\%. We used the results to study the chromatic effects on the depths of exoplanet transits obtained at different epochs and corresponding to different stellar spot distributions. In the case of WASP-52, which has peak-to-peak photometric variations of $\sim$7\% in the visible, the residual effects of dark spots on the measured transit depth of its giant planet, after applying the calculated corrections, are about $10^{-4}$ at 550\,nm and $3\times10^{-5}$ at 6\,$\mu$m.}
{We demonstrate that by using contemporaneous ground-based multiband photometry of an active star, it is possible to reconstruct the parameters and distribution of active regions over time, thus making it feasible to quantify the chromatic effects on the planetary radii measured with transit spectroscopy and mitigate them by about an order of magnitude.
}

          
\maketitle

\section{Introduction} \index{Int} \label{Int}

After the ground-breaking work done during the last decades in the search for exoplanets using both the Doppler technique and photometric transits \citep{1995Natur.378..355M, Charbonneau2000, 2016Natur.536..437A, 2017Natur.542..456G}, the community has expanded its focus to include the characterisation of known transiting planets through the study of their atmospheres. The central idea behind transmission spectroscopy is that the planetary transit depth has a chromatic dependence since its atmosphere selectively absorbs  certain wavelengths. The measured transit depth as a function of wavelength constitutes the planet's transmission spectrum and opens up the possibility of studying the chemical (composition, abundances) and physical (thermal structure, pressure profile) properties of the atmosphere. This has been revealed as a very successful technique \citep{Seager2000, Tinetti2007, Fortney2010, Burrows2014, Sing2015, Tsiaras2019}. The optimal candidates for transmission spectroscopy are low-density planets with atmospheres of low mean molecular weight (hydrogen-dominated) and high equilibrium temperature, thus favouring hot, giant gaseous planets \citep{Stevenson_2016, Kreidberg_2018}. Even for these candidates, the observed transit depth effects induced by the planet's spectral features are the order of $10^{-3}$, which makes their detection very challenging, particularly in the case of active host stars, where other sources of variations are expected.

Future space-based telescopes such as the JWST 
\citep[James Webb Space Telescope;][]{Gardner2006} and the Ariel 
\citep{2018ExA....46..135T} 
missions are designed for this purpose. The latter will carry out a comprehensive, large-scale survey of the chemical compositions and thermal structures of the atmospheres of $\sim$1\,000 known transiting exoplanets in the optical and near-infrared (NIR) wavelengths (0.5 to 8\,$\mu$m) following its expected launch in 2028 \citep{2018ExA....46..135T,2018ExA....46...31E}. 

Magnetically-driven active regions on the surface of a star, such as dark spots and bright faculae, can produce significant alterations on the measured planetary transit depths \citep{Lagrange2010spots, Lagrange2010plages, Barros2013, 2014A&A...568A..99O}. Such inhomogeneities on the stellar photosphere can induce chromatic effects \citep{Sing2015}, which, in the case of unocculted spots, can be very similar to the signature of atmospheric Rayleigh scattering  \citep{2009A&A...494..391R,2014ApJ...791...55M}. \citet{2018ApJ...853..122R} studied their influence in M dwarfs and find it to be up to ten times larger than the effect expected from atmospheric compounds even at NIR wavelengths. Although \citet{2019AJ....157...96R} find the influence of surface phenomena in FGK stars to be generally lower and possibly only measureable for active stars, those studies challenge the reliability of retrieved transmission spectra, at least under certain circumstances. Clouds and hazes can also produce chromatic effects, introducing a Rayleigh-like slope and grey opacity, respectively \citep{Pinhas2017}. Distinguishing the different contributions to wavelength-dependent transit depth variations is therefore of crucial importance, and this calls for sophisticated modelling of photospheric inhomogeneities. 

Several approaches to model the effects of stellar activity on time-series spectrophotometry have been presented. Examples are the SOAP \citep{2012A&A...545A.109B} and SOAP 2.0 \citep{Dumusque2014} codes, which are conceived to provide estimates of the effects of spots and plages on photometry and radial velocities (RVs) by surface integration of a pixel grid using solar observed cross-correlation functions (CCFs). Similar to this is the maximum entropy regularisation method used by \citet{Lanza2011} to model a unique and stable solution for the flux variations due to discrete elements of active regions. An alternative empirical approach is the FF' method \citep{Aigrain2012}, which uses the flux time series to predict the RV jitter. \citet{Herrero2016} presented the \textsf{StarSim} code as a more sophisticated methodology to simulate light curves in various passbands and spectroscopic time series of active rotating stars taking into account the effects of stellar spots and plages. \textsf{StarSim} is based on surface integration techniques, taking into account the particular properties of each element of a fine stellar surface grid such as the effective temperature, the limb darkening and convection effects.

We describe the new version of \textsf{StarSim} in Sect. \ref{S2}, detailing the designed method to retrieve the properties of a star and its activity by solving the inverse problem with real time-series data. The aim of this study is to obtain an activity model which comprises 1) a stellar surface map and, 2) a set of stellar parameters. This methodology is implemented in a test case using observational multiband photometry of the planet host WASP-52, as shown in Sect. \ref{S0}. In Sect. \ref{S4}, we use the retrieved activity model for WASP-52 to study the different sources of chromatic effects on the transit depths of its known planet WASP-52~b. In Sect.\,\ref{S7} we conclude and discuss the possibilities of this methodology in correcting transit spectroscopy data affected by stellar activity.  

\section{\textsf{StarSim} model} \index{S2} \label{S2}

\subsection{The forward model}

The \textsf{StarSim} modelling code generates the effects of evolving dark spots and bright faculae of a rotating star on photometric, RV and activity indices time-series data of magnetically-active stars. We explain the most important points of the modelling approach below, but we refer the reader to the more detailed description in \citet{Herrero2016}. The code divides the stellar photosphere into a grid of elements, each of which is assigned an effective temperature value depending on its nature and according to the distribution of stellar activity features. We use $T_{\rm{ph}}$ for an immaculate element of the photosphere, $T_{\rm{sp}}$ for those occupied by cool spots, and $T_{\rm{fc}}$ for those corresponding to bright faculae, and assume generally that $T_{\rm{sp}}<T_{\rm{ph}}<T_{\rm{fc}}$. We attribute to each surface element the corresponding theoretical BT-Settl spectrum \citep{Allard2013} generated with the Phoenix code. In this approach the only difference between surface features is its effective temperature (and not other spectral effects related, e.g. to the stellar chromosphere), as this is the parameter that dominates their chromatic signature on the simulated data. Activity features on the photosphere are described by a set of magnetically-active elements. Each active element is represented with a circular spot that can be surrounded by a concentric facular area. The model can take into account arbitrary shapes of the active regions by grouping several smaller active elements together. Throughout the text, we indistinctly use the word spot or the expression active element.

As a simplification aimed at a faster execution of the code, the properties of the stellar surface, such as the limb darkening, the radial velocity, and the projection angle towards the observer, are assumed to be constant inside each active element. Therefore, active elements are always assumed to be small as they are associated with the properties of a single pixel on the surface grid. In this version of \textsf{StarSim} we employ active elements having circular shape. Besides the different temperature, a set of six parameters are used to characterise an active region: the time of appearance of the active element, its lifetime, its location on the surface of the star (latitude and longitude), its angular radius, and the facula-to-spot area ratio $Q$. The last one is the only parameter assumed to be the same for all active elements. The model includes all the relevant geometric, physical and wavelength-dependent features related to the presence of active regions, such as limb darkening effects computed from Kurucz ATLAS9 models \citep{2017ascl.soft10017K}, limb brightening in the case of faculae \citep{2010A&A...512A..39M}, convection effects inside and outside the active regions \citep[see][for a detailed description of the convection model]{Herrero2016}, and the projection of each surface element towards the observer.

Photometric and spectroscopic time-series are obtained by integrating over all the grid elements at each time step, taking into account the rotation of the star and the time-evolution of the active regions. When simulating photometric time-series, the integrated spectrum of the projected stellar surface is computed at each time step, and then multiplied by the defined filter passbands in order to retrieve multiband fluxes and produce light curves. By construction, the resulting light curves are relative to the immaculate photosphere (i.e. immaculate flux = 1). To compare with observational measurements, a zero-point correction needs to be applied to each band. This is done by normalizing both the observed light curve and the model light curve to the weighted average of all measurements. In the case of the simulated light curve, only the epochs with observational measurements are used. For the calculation of RVs and other spectral indices related to CCFs, \textsf{StarSim} initially generates the CCFs produced from the spectrum of a single photosphere, spot and facular element, and then integrates the entire visible surface using CCFs instead of spectra. This is necessary to speed up the execution of the code and make it possible to handle with normal desktop computers. A slow rotator template or a user-defined mask of spectral lines can be used to calculate the CCFs. In a first step of each time-series simulation, the contribution of the entire immaculate photosphere is computed without any active elements. Then, for each time step of the simulated data series, the contribution of the active regions is added by considering their individual location on the observed stellar disc.

\subsection{Analytical foundations of the inverse problem} \label{foundations}

\begin{figure}
\begin{center}
\includegraphics[width=6.0cm]{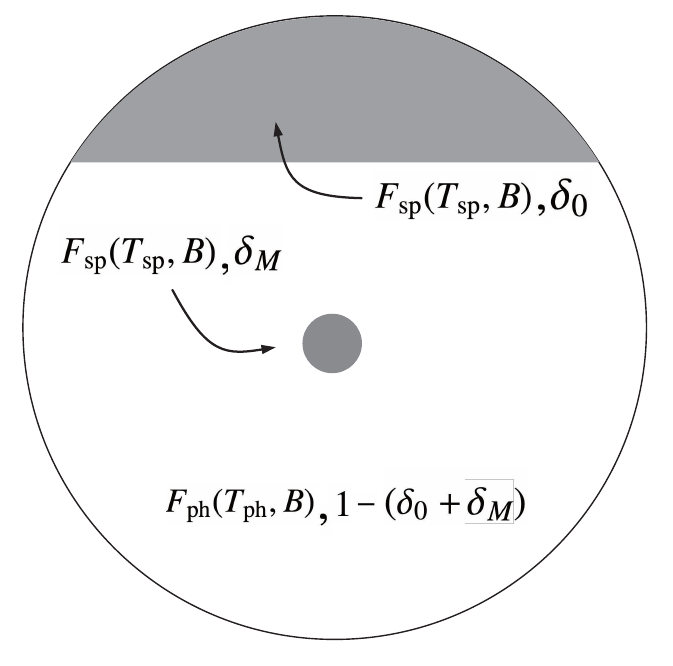}
\end{center}
\caption{Example of a stellar photosphere including modulating and non-modulating spots. The model is composed of a polar cap, which does not contribute to light curve modulation, with a projected filling factor $\delta_0$, and an equatorial small spot with a maximum projected filling factor $\delta_M$. $F(\cdot)$ is the brightness of each surface element. }        
\label{fig:Spotted_sketch}
\end{figure}

Before moving on to discuss how \textsf{StarSim} handles the inversion of photometric light curves, we shall present the analytical foundations of the model and the relevant variables. We consider a stellar surface with active regions covering a total projected filling factor $\delta_{\rm sp}$, which is made of two components: one that produces flux variations over time due to rotational modulation, $\delta_M$, and another one that stays constant as the star rotates (e.g. a polar cap or a uniformly spotted latitudinal band), $\delta_0$. To demonstrate analytically that light curves can carry information of both spottedness levels we assume a simple model consisting of a star with a circular spot on its equator, with a surface $\delta_M$, and a polar cap covering a surface $\delta_0$, as illustrated in Fig.\,\ref{fig:Spotted_sketch}.

The amplitude of the photometric modulation can be estimated as the ratio of the flux of the star when the modulating spot is out of view,
\begin{equation}
\label{eq:f1}
f_1 = F_{\rm{sp}}(\mathcal{B}, T_{\rm sp})\,\delta_0 + F_{\rm{ph}}(\mathcal{B}, T_{\rm ph})\,(1-\delta_0),
\end{equation}
with respect to the phase when the spot is at the centre of the visible stellar disc,
\begin{equation}
\label{eq:f2}
f_2 = F_{\rm{sp}}(\mathcal{B}, T_{\rm sp})\,(\delta_0 + \delta_M) + F_{\rm{ph}}(\mathcal{B}, T_{\rm ph})\,(1-\delta_0-\delta_M),    
\end{equation}
where $F_{\rm{ph}}$ and $F_{\rm{sp}}$ are the surface fluxes of the spotted and immaculate surface of the star, respectively, which depend on the temperature of each surface element ($T_{\rm ph}$ and $T_{\rm sp}$) and the spectral band ($\mathcal{B}$). The relative amplitude of the photometric variability can be computed as $(f_{1}-f_{2})/f_{1}$, which is equivalent to the measurement provided by observational light curves. Rearranging Eqs.\,(\ref{eq:f1}) and (\ref{eq:f2}), the photometric amplitude can be written as,
\begin{equation}
\label{eq:amplitude}
A_{\rm obs}(\mathcal{B}, \Delta T_{\rm sp}, \delta_M, \delta_0) = \frac{\delta_M}{ \Phi(\mathcal{B},\Delta T_{\rm sp}) - \delta_{0}},
\end{equation}
where
\begin{equation}
\label{eq:bandfactor}
\Phi(\mathcal{B},\Delta T_{\rm sp})= \frac{1}{1-\Big(\frac{F_{\rm sp}}{F_{\rm  ph}}\Big)_{\mathcal{B},\Delta T_{\rm sp}}},
\end{equation}
with $\Delta T_{\rm sp}$ being the difference between the photospheric and spot temperature ($\Delta T_{\rm sp} = T_{\rm ph} - T_{\rm sp}$). We assume  that the effective temperature of the star can be estimated independently. These equations illustrate that the photometric amplitude increases linearly with the modulating filling factor $\delta_M$, and decreases with a combination of the non-modulating filling factor $\delta_0$ and the brightness contrast between the photosphere and the spots $\Phi$, which is, in turn, a function of the photometric band ($\mathcal{B}$) and the temperature contrast ($\Delta T_{\rm sp}$). We remark here that these quantities satisfy some constraints, namely, $0 \leq \delta_0 \leq 1$, $0 \leq \delta_M \leq 1$, $0 \leq \delta_0 + \delta_M \leq 1$ and $\Phi>1$, for cool spots.

\begin{figure}
\begin{center}
\includegraphics[width=7.7cm]{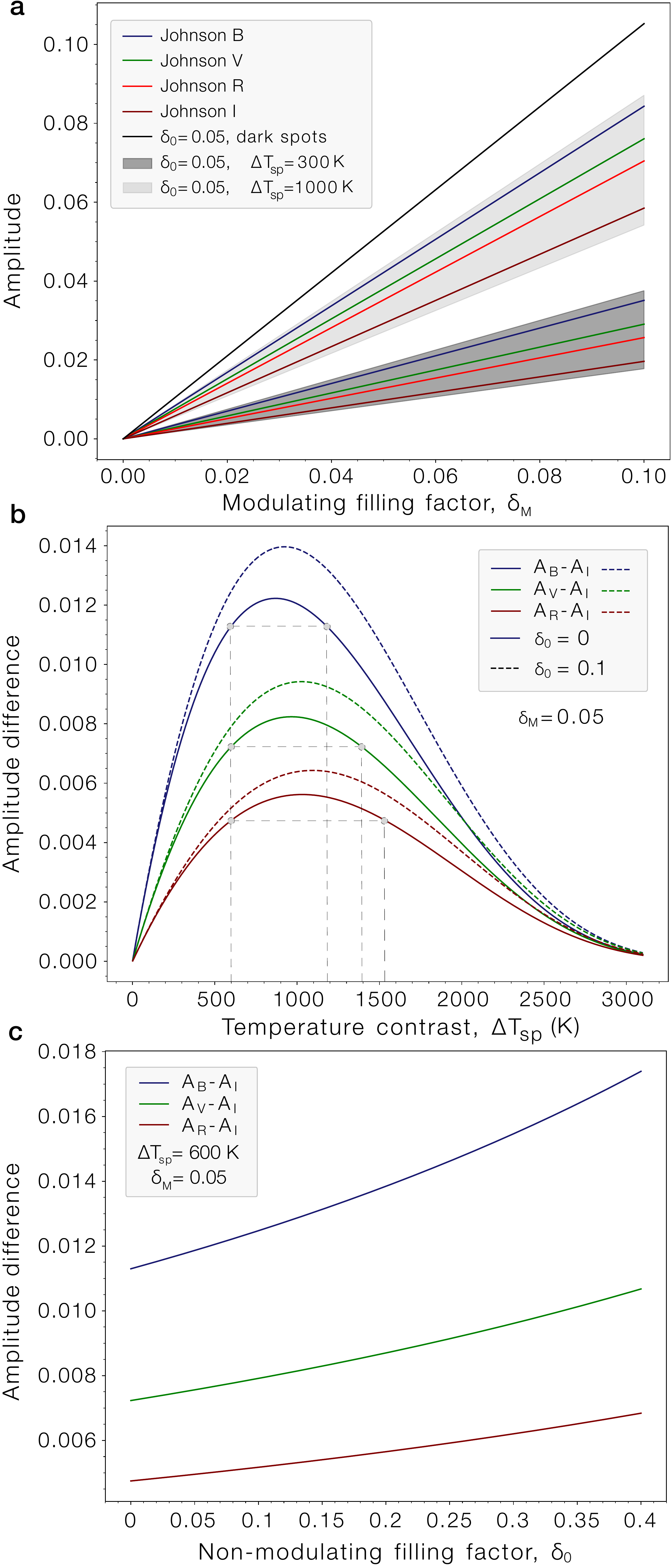}
\end{center}
\caption{Graphical representation of Eq.\,(\ref{eq:amplitude}) for the parameters of interest assuming a stellar photospheric temperature of $T_{\rm ph}=5000$\,K and a black-body for the spectral energy distribution of each surface element. \textbf{a)} Photometric amplitude as a function of the modulating filling factor for different spot temperature contrasts. A non-modulating filling factor $\delta_0=0.05$ is employed, but adopting other values has negligible effect at this scale. The gray area covers the spectral range from $400$\,nm to 1\,$\mu$m. \textbf{b)} Difference in activity-induced amplitude for photometric bands $BVR$ with respect to the $I$ band as a function of spot temperature contrast. A value of 0.05 is adopted as modulating filling factor ($\delta_M$) and examples for non-modulating filling factors of $\delta_0=0$ and 0.1 are shown. The gray lines illustrate how the bi-valuate nature of the temperature contrast effect can be resolved if more than two bands are used. \textbf{c)} Difference in activity-induced amplitude for photometric bands $BVR$ with respect to the $I$ band as a function of the non-modulating filling factor. A modulating filling factor $\delta_M=0.05$ and a spot temperature contrast $\Delta T_{\rm sp} = 600$\,K are adopted.}
\label{fig:fillings}
\end{figure}

Figure\,\ref{fig:fillings} illustrates Eq.\,(\ref{eq:amplitude}) and its dependence on the three independent parameters. If only information from a single band is available (e.g. typical data from exoplanet surveys), the linear dependence of the light curve amplitude with $\delta_M$ permits the determination of the modulating filling factor provided the non-modulating filling factor is neglected and the spot temperature contrast is adopted as an external constraint. This has been a common practice in the literature. In case two or more bands are available, the possibility of determining another variable such as $\Delta T_{\rm sp}$ or $\delta_0$ arises. For typical spot temperature contrasts and visible bands, it can be shown that $\Phi$ is significantly greater than $\delta_0$ and therefore the former parameter dominates. Thus, from two or more photometric bands (preferably covering a large interval in wavelength, i.e. large $\Phi$ variation), the simultaneous determination of $\delta_M$ and $\Delta T_{\rm sp}$ becomes possible. A practical application can be seen in \cite{Mallonn_2018}. In the particular case of two bands, the solution becomes bi-valuate, with two possible $\Delta T_{\rm sp}$ reproducing the amplitude difference, as shown in Fig.\,\ref{fig:fillings}b. However, if a third band is available, such degeneracy can be broken, as illustrated by the gray lines in the figure. Eq.\,(\ref{eq:amplitude}) further shows that, from three bands or more, one can theoretically determine at the same time the three relevant variables. For that to be possible, the photometric information needs to be of sufficient precision to discriminate the changes induced by each variable. The curvature in Fig. \ref{fig:fillings}c (i.e. the variation in the amplitude difference) is what makes it possible to determine the non-modulating filling factor from multi-colour photometry. Nevertheless, the scale of the variations makes reliable estimates of $\delta_0$ very challenging for typical ground-based photometric precisions.

We remark that we have presented a simplified version of the problem, defined by only two epochs (maximum and minimum light). However, photometric time series also carry information on the relevant parameters because of the correlations present among the different measurement epochs, thus adding additional constraints to the spot properties. The formulation discussed here proves that it is possible to simultaneously determine the total spot filling factor and spot temperature contrast as long as good multiband photometric data are available, therefore providing theoretical foundation to the inverse problem.

\subsection{The inverse problem}
\label{sec:inverse_problem}

The most recent \textsf{StarSim} version can perform the inverse problem. The goal is to obtain the underlying properties, that is to say a stellar activity model as described by the parameters of the star and its surface map, that reproduce the observed time-series data $\mathbf{X}$. This type of non-linear problem can be expressed as
\begin{equation}
\mathbf{X} = \textsf{F}(\mathcal{S}, \theta) + \epsilon,
\label{eq:inverse_problem} 
\end{equation}
where $\mathbf{X}$ is the time-series data, $\textsf{F}$ is the activity model, and $\theta$ is a set of stellar parameters. The surface map $\mathcal{S}$ is the set of parameters that describe the surface distribution, sizes, and lifetimes of all the active elements considered, each of them defined as a small circular spot surrounded by a bright facular region. Finally, $\epsilon$ is an additional noise term, or jitter, that we adopt as uncorrelated and following a Gaussian distribution (white noise). 

The inverse problem consists in finding the surface map that best reproduces the data $\textbf{X}$ for a given $\theta$, expressed as
\begin{equation}
\hat{\mathcal{S}}_{\theta} = \textsf{F}^{-1}(\textbf{X}, \theta), 
\label{eq:inverse_problem_2} 
\end{equation}
\noindent
where $\hat{\mathcal{S}}_{\theta}$ means the optimal surface map linked to a specific set of stellar parameters. Subsequently, this map and its associated parameters provide an optimal fit to the photometric and spectroscopic time-series data when applying the forward model (Eq.\,\ref{eq:inverse_problem}).

\subsubsection{Objective function}

The statistical function to optimise, or figure of merit, is a linear combination of the logarithmic likelihood function of all the time-series data defined as
\begin{equation}
\ln \mathcal{L}\,(\mathcal{S},\boldsymbol{\theta}) = \sum_{j}^{\textrm{\textsf{\tiny{Obs.}}}} a_j\,\ln \mathcal{L}_j(\mathbf{X}_j \, | \, \mathcal{M}_j(\mathcal{S}, \boldsymbol{\theta})),    
\label{eq:joint_statistic} 
\end{equation}
\noindent
where $\ln \mathcal{L}_{j}$ is the log-likelihood of the fitted model $\mathcal{M}_j$ according to the observational data $\mathbf{X}_j$ for the $j$-th observable. The quantities $a_j$ are the weights associated to each set of observables (here we assume $a_j=1$). As shown in Eq.\,(\ref{eq:inverse_problem}), we consider the simplest case of non-correlated Gaussian uncertainties (white noise). The likelihood function is then written as
\begin{equation}
\mathcal{L}_j = \prod_{i}^{N} \frac{1}{\sqrt{2\pi}(\sigma_i^2+s_j^2)} \exp \Big[ -\frac{(y_i-\mathcal{M}_{ij})^2}{2(\sigma_i^2+s_j^2)} \Big],
\label{eqn:versembl} 
\end{equation}
\noindent
where $\mathcal{M}_j$ is the \textsf{StarSim}-generated model of the $j$-th time-series observable with $N$ measurements, and $y_i$ are the observational data points. $\sigma_i$ is the nominal error of the measurement $i$, and $s_j$ is a quadrature-added jitter, that accounts for a possible incompleteness of the model, underestimated uncertainties or traces of correlated noise.
 
\subsubsection{Optimizing the surface distribution of active regions}
\label{sec:gen_spot_maps}

Surface maps describing the distribution, size and evolution of the active elements are obtained through maximisation of the figure of merit (Eq.\,\ref{eq:joint_statistic}) given a fixed set of stellar parameters, $\theta$, typically including the rotation period, $P_{\rm rot}$, the spot temperature contrast, $\Delta T_{\rm sp}$, and the facula-to-spot area ratio, $Q$. Thus, the optimal surface map is given by
\begin{equation}
 \hat{\mathcal{S}}_{\theta} = \arg \max_{\mathcal{S}_j \in \boldsymbol{\mathcal{S}}} \ln \mathcal{L}_{\theta}(\mathcal{S}) ,
\label{eq:argmax} 
\end{equation} 
\noindent
where $\hat{\mathcal{S}}_{\theta}$ is the optimal surface map constrained to a specific set of parameters and $\mathcal{S}_j$ is any surface map among all possible maps, $\boldsymbol{\mathcal{S}}$.   
In order to optimise the expression in Eq. (\ref{eq:argmax}), we implemented a Monte Carlo Simulated Annealing optimisation algorithm \citep[hereafter MCSA,][]{Kirkpatrick1983}.  

The surface map optimisation starts with a random distribution of a fixed number of active elements, each one characterised by 5 adjustable parameters: time of appearance, lifetime, latitude, longitude, and angular radius. Active elements do not appear or disappear abruptly. They are assumed to grow or shrink linearly in radius at a rate of $0.5$ deg/day \citep[see][for additional details]{Herrero2016}. The value of $Q$ is assumed to be the same for all spot elements. The total number of active regions is a parameter of \textsf{StarSim} and can be defined by the user. However, its is advisable to limit the maximum number of active elements thus forcing the model to retrieve simpler surface maps, avoiding the appearance of a large number of very small spots, especially when modelling noisy time-series data.

The optimisation process follows successive iterations where the algorithm randomly selects one of the adjustable parameters from a randomly-selected active element and modifies it slightly. Then, with this new map, the forward problem is re-computed and $\ln\mathcal{L}$ is evaluated for each time-series dataset. If such perturbed map improves the quality of the fit as given by Eq.\,(\ref{eq:joint_statistic}), the change is accepted. Otherwise, it is accepted with a probability $e^{-\beta\Delta\ln\mathcal{L}}$, where $\beta$ is a parameter that grows in each step. With this strategy, the optimiser avoids getting trapped in local maxima. A more detailed explanation of the implementation of MCSA in \textsf{StarSim} can be found in \citet{Herrero2016}.

Due to the large number of parameters describing the stellar surface map and the intrinsic randomness of Monte Carlo algorithms, several maps retrieved with the same $\theta$ may not be identical in spite of producing very similar time-series datasets with the forward model. This effect can be mitigated by performing a number of solutions with different initial spot maps and subsequently exploring the variance of the likelihood statistic. For each target and set of observations, these tests can help the user to define parameters such as the number of iterations of the MCSA algorithm per $\beta$ step and the number of active regions on the stellar surface. The execution time (which increases linearly with both parameters) and some regularisation criteria (selecting the minimum number of spots to avoid unnecessarily complex surface structures) need to be factored in when deciding on the optimal procedure. Once these parameters are adopted and a large number of inversions with varying initial spot conditions are run, the final product of the inversion for a fixed set of parameters, $\theta$, is a spot map calculated by averaging the total sample of optimal maps. The result is a smooth time-evolving surface map that contains valuable information about the surface distribution of the active regions, their typical lifetimes and possible differential rotation.

\subsubsection{Optimizing stellar parameters}
\label{subsec:optimal_theta}

Equation (\ref{eq:argmax}) describes the optimisation of the surface maps $\mathcal{S}_{\theta}$ when fixing a set of stellar parameters $\theta$. However, tests show that $\mathcal{S}$ has a dependence on $\theta$. Due to the non-linear and multimodal nature of the problem, $\mathcal{S}_{\theta}$ and $\theta$ are strongly coupled, and small variations on the parameters potentially imply large changes in $\mathcal{S}$. Therefore, fitting both sets of parameters simultaneously is computationally challenging. Therefore, we choose a two-step approach. The retrieval of optimal parameters in this context belong to the class of noisy optimisation problems \citep[e.g.][]{noisy_opt}, which are characterised by long evaluation times and noisy outputs for the objective function. In our particular case, we consider a large number of randomly-generated parameter sets drawn according to a prior distribution (generally uniform) and calculate the inverse problem to produce an optimal surface map starting from random initial conditions for each of the parameter sets. Using their $\ln \mathcal{L}$ values, a certain interval enclosing the best statistically-equivalent solutions can be defined. From the selected best solutions, the optimal parameter set and the corresponding uncertainties can be determined. The procedure can be expressed as
\begin{multline}
\{\theta\}^* = \{\theta_{\ln \mathcal{L}_0},\theta_{\ln \mathcal{L}_1}, \ldots, \theta_{\ln \mathcal{L}_{\rm N}} | \ln \mathcal{L}_j \geq \ln \mathcal{L}_{j+1}, \Delta \ln \mathcal{L}_{\rm MAX} = \\
= \ln \mathcal{L}_{0}-\ln \mathcal{L}_{\rm N} \sim \Delta \ln \mathcal{L}_{\rm lim}\},
\label{eq:theta_sorted_list} 
\end{multline} 
\noindent
where $\Delta \ln \mathcal{L}_{\rm lim}$ is the threshold defining equivalent solutions. Then, the median and 68\% confidence interval of $\{\theta\}^*$ are used as the best estimates of stellar parameters and their uncertainties.

\begin{figure}
\begin{center}
\includegraphics[width=8.5cm]{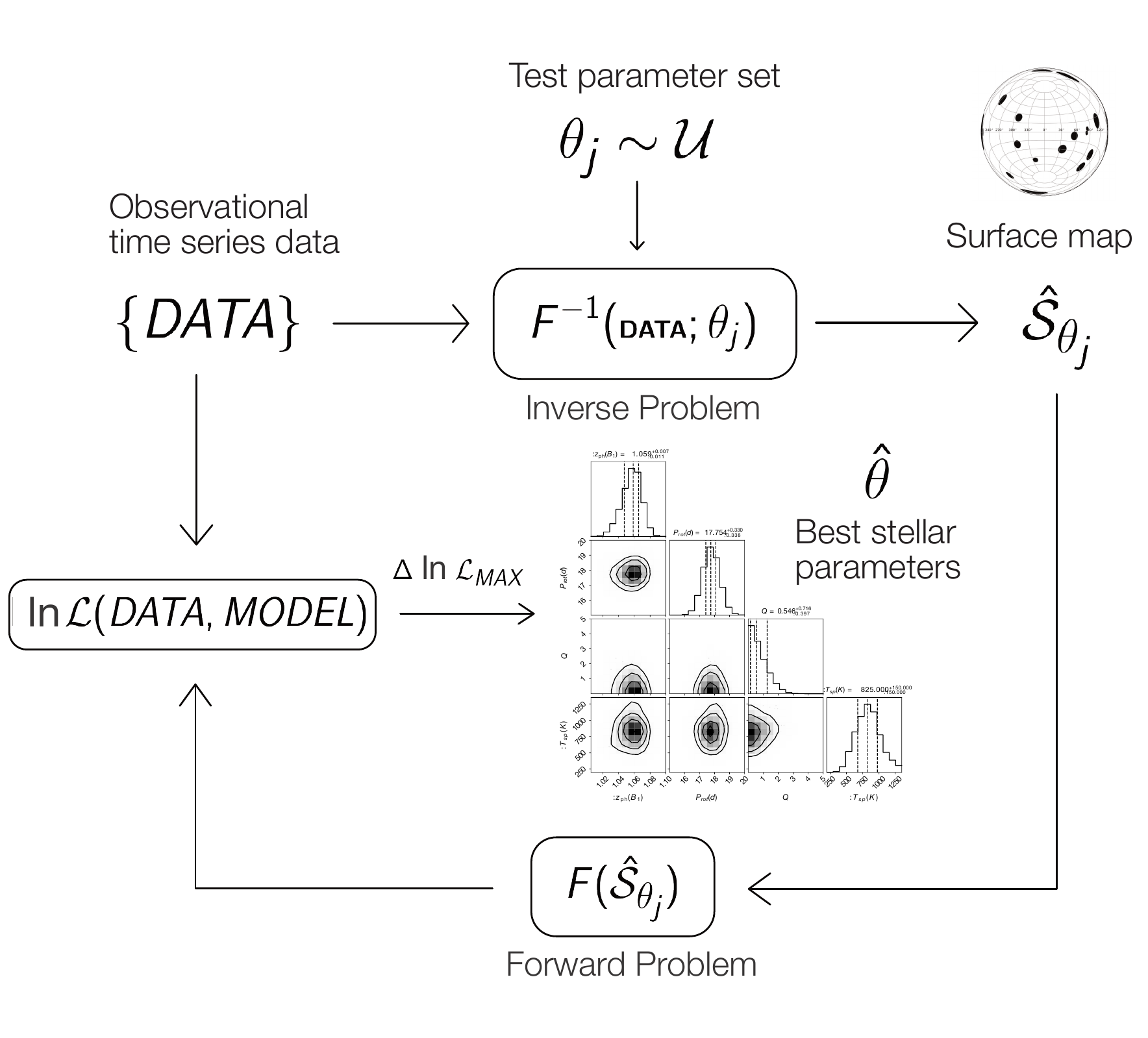}
\end{center}
\caption{Flow chart of the parameter retrieval approach as presented in Sect. \ref{sec:inverse_problem}. 
}        
\label{fig:flow_chart}
\end{figure}

\subsubsection{Degeneracies of the inverse problem}
\index{S4x}
\label{sec:optimization}

The methodology of \textsf{StarSim} to calculate the inverse problem implies accounting for a large number of parameters and potential correlations among them, which may produce degenerate solutions. This is especially important when only a particular type of data is considered (i.e. photometry or spectroscopy). A number of different active region configurations can yield very similar simulated datasets, with figures of merit that are not significantly different. This can be easily illustrated by the example of considering a system with a single spot on an equator-on star ($i_{\rm{star}} = 90^{\circ}$). In such case, the location in either hemisphere produces identical results for all datasets (hemispherical degeneracy). Also, spots with different sizes at different latitudes can produce similar effects, as they may yield the same projected area towards the observer. This results in size-latitude correlations. Such type of weak degeneracies can be solved with an accurate modelling of limb darkening. However, the uncertainty of photometric and spectroscopic measurements usually make limb darkening variations indistinguishable. This is why, especially for stars that are nearly equator-on, we cannot retrieve the latitude distribution of the active regions but only the stellar longitudes occupied by spots. Therefore, surface maps can only be shown as the filling factor of active regions projected on the longitude axis (see Sect.~\ref{surface_map} for the case study presented here). Also, when polar and circumpolar active regions are present, their integrated effect can result in a constant offset of the photometric observables, which we have shown that can be difficult to identify. The inverse problem in case of stars that are not equator-on is particularly challenging when only photometry is available. 

Another important potential degeneracy is caused by the correlation between the size of the active regions and their temperature contrast. However, there is potential relief to such degeneracy because the temperature contrast of spots and faculae introduces a chromatic effect, as we have shown above. This can be measurable when multiband time-series photometry is available. When cool spots are present on the stellar surface, the amplitude of the photometric variability is larger at bluer wavelengths, and lower in the red and infrared bands. Such chromatic signature can be studied by performing the inverse problem with \textsf{StarSim} on multiband photometric measurements, thus allowing to derive the temperature contrast of the active regions with respect to the photosphere and thus break the temperature-size correlation. 

Finally, different combinations of the filling factor of spots and the amount of faculae ($Q$) can also result in similar solutions for the modelled observables. This is because bright regions can partially compensate the effects of dark spots. However, the presence of faculae can also be identified when analyzing multiband photometric time series, with proper modelling of limb brightening (as opposed to limb darkening for dark spots) and the corresponding chromatic signature, which differs from the one produced by cool spots. Also, due to the blocking of convection by active regions and the subsequent decrease of the net convective blueshift in RVs, the spot-facula degeneracy can also be broken when contemporaneous photometry and high-precision spectroscopy are available.

\subsubsection{Toy model inversion}

As a further test of the validity of our approach, we conducted an inversion exercise using data generated with the forward functionality of \textsf{StarSim}. We considered a star with $T_{\rm ph}=5000$~K, $i=90$ deg, $P_{\rm rot}=15$ days and with its surface covered by 5 spots having a $\Delta T_{\rm sp} = 700$~K and no faculae ($Q=0$). We considered spots of various sizes, placing them at different latitudes and longitudes in such a way that a photometric modulation is present and also ensuring that spotted regions are visible at all times (i.e. projected spot filling factor never being zero), to produce a nonzero non-modulating filling factor. We assumed the spots to be of constant size and static in the frame of reference of the rotating star. We generated synthetic light curves in the $UBVRIJ$ bands covering a timespan of 90 days (6 rotations) and with a uniform observational cadence of one measurement every 0.5 days. Noise was added to the photometric measurements following a heteroskedastic scheme, consisting of non-correlated Gaussian noise $\sim\mathcal{N}(0,\sigma)$ with $\sigma\sim\mathcal{U}(0.0085,0.00115)$ (i.e. $\approx$1000 ppm).

\begin{figure*}
\begin{center}
\includegraphics[width=18cm]{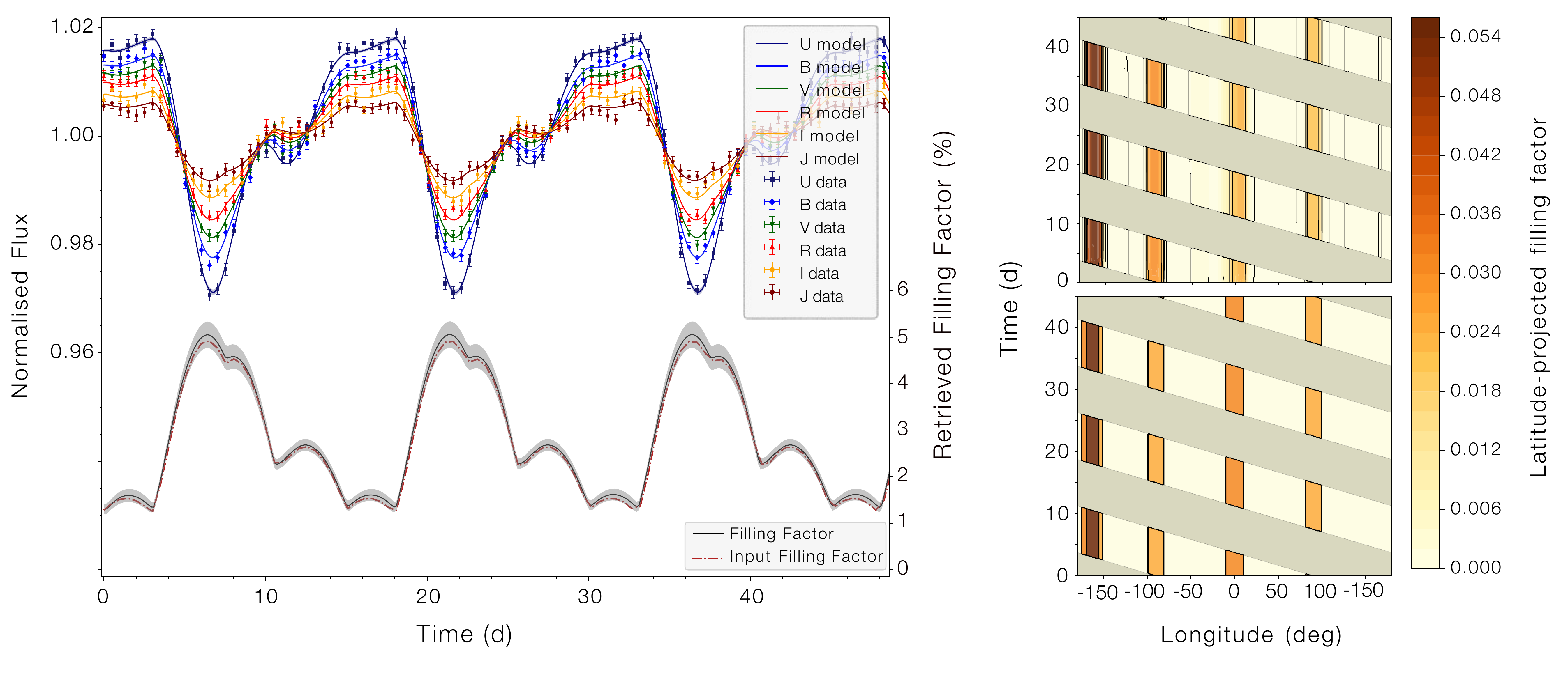}
\end{center}
\caption{Left panel shows the \textsf{StarSim} model fits to multiband synthetic light curves assuming the parameters in Table\,\ref{tab:toy_model_params}, for a subset of 50 days. Solid curves represent the mean of $\sim$20 optimal solutions of the inverse problem. The gray line at the bottom is the projected spot filling factor of the maps, also showing the mean and 1-$\sigma$ shaded band. The error intervals of the photometry bands are very small and difficult to recognise. The right panels plot the longitudinal spot filling factor projected on the stellar equator as a function of time, for both the input (bottom) and the retrieved (top) spot maps. Only stellar longitudes visible at each time are shown (hence the band structure). The colour scale indicates the fractional latitude-projected spot coverage for each 1-degree longitude bin. 
}        
\label{fig:toy_model}
\end{figure*}

The resulting 6 light curves were inverted following the scheme sketched in Fig.\,\ref{fig:flow_chart}. The fitted parameters were $P_{\rm rot}$, $Q$, and $\Delta T_{\rm sp}$. We ran several thousand inversions exploring the relevant parameter space, all starting from a random map. We considered three different assumptions regarding the number of active regions in the model, namely 5, 10, and 20, to evaluate the impact on the quality of the solution and on the resulting filling factor. We furthermore considered that spots have a finite lifetime and that their growth or decrease rate is $0.5$ deg$\times$ day $^{-1}$. Each spot was characterised by six parameters (appearance time, lifetime, longitude, latitude, radius) that were allowed to vary during the optimisation process. We established a likelihood criterion to select the best maps and solutions based on the inherent scatter of the solution when performing a large number of inversions from random maps and assuming the correct parameters. This yielded a number of $100-150$ good solutions for all three cases. The statistical results of such solutions are given in Table\,\ref{tab:toy_model_params}, and a graphical illustration for the 10-spot case is shown in Fig.\,\ref{fig:toy_model}.

The results of the inversion tests allow assessing the retrieval performance of the \textsf{StarSim} model. As can be seen in Table\,\ref{tab:toy_model_params}, all fitted parameters are retrieved within the uncertainties regardless of the number of spots assumed. No significant bias is observed except for a slight tendency to underestimate the spot temperature contrast. However, additional tests showed that this may be caused by the relatively low number of solutions used and therefore it should not be reason for concern. It is interesting to note that the spot filling factor is also accurately obtained, including the non-modulating fraction. This indicates that the algorithm is able to reduce the size of the spots and place them appropriately to reproduce the correct spottedness of the star. We do not see a dependence on the number of active regions except for a slight tendency to overestimate the filling factor when 20 spots are used. The graphics in Fig.\,\ref{fig:toy_model} show the high quality of the multiband fits and the low dispersion of the solutions. Also, the comparison of the input latitude-projected filling factor with the inversion results on the right panels of the figure indicates a very precise retrieval of both the longitudes and sizes of the active regions. We additionally ran tests using only the $BVRI$ passbands, thus restricting the wavelength baseline. The solutions converged well only with slightly larger uncertainties, as expected. In all cases, the input parameters were well within the error bars.

\begin{table}
\centering
\small
\caption{Results of inversion tests using synthetic data with \textsf{StarSim}. $\delta_{\rm sp}$ is the projected filling factor.}
\label{tab:toy_model_params}
\begin{tabular}{l l c c c}
\hline \hline 
  Parameter & Input &  \multicolumn{3}{c}{\emph{Model}} \\ 
   &  & $5$ spots & $10$ spots & $20$ spots \\ 
\hline  
\noalign{\smallskip}
  $P_{\rm rot}$(d) & $15$ & $15.001^{+0.011}_{-0.011}$ & $15.002^{+0.011}_{-0.013}$ & $14.995^{+0.015}_{-0.012}$\\
   \noalign{\smallskip}
  $\Delta T_{\rm sp}$(K)  & $700$ & $650^{+100}_{-125}$ & $700^{+50}_{-75}$ & $675^{+50}_{-50}$ \\
 \noalign{\smallskip}
 $Q$ &  $0$ & $<0.23$ & $<0.22$ & $<0.13$ \\
 \noalign{\smallskip}
 $\delta_{\rm sp}$ (max) & $0.049$ & $0.048^{+0.004}_{-0.004}$ & $0.051^{+0.003}_{-0.003}$ & $0.053^{+0.004}_{-0.004}$ \\
 \noalign{\smallskip}
 $\delta_{\rm sp}$ (min) & $0.012$ & $0.012^{+0.002}_{-0.002}$ & $0.013^{+0.001}_{-0.001}$ & $0.015^{+0.002}_{-0.002}$ \\
 \noalign{\smallskip}
\hline \hline
\end{tabular}
\end{table}

\section{An activity model for WASP-52} \index{S0} \label{S0}

As an example of the use of \textsf{StarSim} for the inverse problem, we use multiband photometric time-series data of the active planet host star WASP-52 to retrieve an optimal set of stellar parameters and reconstruct a time-variable probability map of the distribution of active regions.

\subsection{The WASP-52 exoplanetary system} 

WASP-52 is a young 
and active K2\,V star hosting an inflated Jupiter-sized exoplanet with an orbital period of 1.75\,days \citep{2013A&A...549A.134H}. The inclination of the planetary orbit $i$ is such that results in a transit of the planet.
\citet{2013A&A...549A.134H, Louden_2017, Mancini2017, May_2018, 2018A&A...619A.150O, Ozturk_2019} described and refined the different planetary parameters using light curves and the Rossiter-McLaughlin effect (see Table\,\ref{tab:WASPparams}), whereas \citet{Kirk_2016, Louden_2017, Chen_2017, 2018AJ....156..298A} described its cloudy sodium-bearing atmosphere. In many of those studies, in-transit anomalies appearing on the transit light curves and the effects of active regions on the stellar photosphere are discussed. \citet{May_2018} and \cite{Bruno_2018} quantify the differences of the chromatic effect on transit depths from a spotted and unspotted photosphere. The different values found for the stellar rotation period $P_{\mathrm{rot}}$ and the spot temperature differences $\Delta T_{\mathrm{sp}}$, which are fitted parameters of our inverse problem, are given in Table\,\ref{tab:simparams_all2}.

In \citet{Alam_2018}, a transmission spectroscopy study was done using three transits observed in the visible and NIR wavelengths with HST/STIS, combined with Spitzer/IRAC photometry. A stellar activity correction was applied by fitting the baseline flux from ground-based photometry using a quasi-periodic Gaussian process. The effective temperature of the star spots was assumed to be 4750\,K, that is, 250\,K cooler than the photosphere. A recent atmosphere study of WASP-52b by \cite{Bruno_2019} combine the STIS and IRAC data from \citet{Alam_2018} and HST/WFC3 from \citet{Bruno_2018}. The spot effects were corrected using the approach described in \citet{Rackham_2017} for an heterogeneous photosphere. In both cases, the stellar photosphere was assumed to be dominated by cool spots.  It is worth emphasising that \cite{Bruno_2019} present a joint fit of the atmospheric model and a stellar contamination correction, parameterised by the fraction of stellar surface occupied by activity features and their temperature. Such approach may lead to a biased solution since genuine planetary features could be modelled as stellar contamination. 

\begin{table}
\centering
\caption{Important parameters of WASP\,52 (top) and WASP\,52b (bottom).}
\label{tab:WASPparams}
\begin{tabular}{l c c c}
\hline \hline \noalign{\smallskip}
parameter & unit & value & ref. \\
\noalign{\smallskip} \hline \noalign{\smallskip}
$\alpha$  & J2000 & 23:13:58.75 & Ga18 \\ 
$\delta$ & J2000 & +08:45:39.9 & Ga18 \\ 
Sp.type & - & K2\,V     & He13 \\
$G$ & mag & 11.954$\pm$0.001 & Ga18 \\ 
$\mu_{\alpha}$ & mas\,a$^{-1}$ & $-$6.914$\pm$0.079 & Ga18 \\
$\mu_{\delta}$ & mas\,a$^{-1}$ & $-$44.248$\pm$0.054 & Ga18 \\
Distance & pc & 175.7$\pm$1.3 & Ga18 \\
M$_*$ & M$_{\odot}$ & 0.81$\pm$0.05 & Ma17 \\
R$_*$ & R$_{\odot}$ & 0.860$^{+0.021}_{-0.027}$ & Ga18 \\
L$_*$ & L$_{\odot}$ & 0.4189$\pm$0.0046 & Ga18 \\
T$_{\rm eff.}^{a}$ & K & 5010$^{+80}_{-60}$ & Ga18 \\ 
$\log{\rm g}^{a}$ & cgs & 4.582$\pm$0.014 & He13 \\ 
age  & Ga & 0.4$^{+0.3}_{-0.2}$  & He13 \\
$\log {R}'_{\rm HK}$ & cgs & $-$4.4$\pm$0.2 & He13\\
$\lambda$ & deg &  5.4$^{4.6}_{-4.2}$ & {\"O}z18 \\
\noalign{\smallskip} \hline \noalign{\smallskip}
M$_{\rm p}$ & M$_{\rm J}$ & 0.46$\pm$0.02 & He13 \\
R$_{\rm p}$ & R$_{\rm J}$ & 1.223$\pm$0.062 & {\"O}z19 \\
P$^{b}$  & days & 1.7497828$\pm$0.0000006 & {\"O}z19 \\ 
T$_{0}^{b}$  & BJD & 2,405,793.68128$\pm$0.00049 & {\"O}z19 \\  
$r/R_*^{b}$ & - & 0.159$\pm$0.004  & {\"O}z19 \\  
$b/R_*^{b}$ & - & 0.60$\pm$0.02 & He13 \\  
Orbit inc., i$^{b}$  &  deg  & 85.24$\pm$0.84 & {\"O}z19 \\ 
a  & au  & 0.0272$\pm$0.0003 & He13 \\
\noalign{\smallskip} \hline \hline
\end{tabular}
\tablefoot{
(a) fixed parameters in Sect. \ref{inver52}; (b) fixed parameters in Sect. \ref{S4}.\\
{\bf References.} He13: \citet{2013A&A...549A.134H};
Ki16: \citet{Kirk_2016};
Ma17: \citet{Mancini2017};
Os18: \citet{2018A&A...619A.150O};
Ga18: \citet{2018A&A...616A...1G};
{\"O}z19 \citep{Ozturk_2019} \\
}
\end{table}

\subsection{Photometric observations} \index{S1} \label{S1}

Photometric light curves obtained from two different ground-based observatories, the \mbox{1.2-m} STELLA telescope at Iza\~na Observatory in Tenerife and the \mbox{0.8-m} Joan Or\'o telescope (TJO) at the Montsec Astronomical Observatory in Catalonia, were used in this work. WASP-52 was observed as part of our monitoring survey VAriability MOnitoring of exoplanet host Stars \citep[VAMOS,][]{Mallonn_2018}. Measurements cover a time interval of more than 500\,days in two different observing seasons (2016 and 2017), and were obtained contemporaneously with both telescopes using four different passbands ($BVRI$). 

The STELLA telescope and its wide-field imager WiFSIP \citep{2004AN....325..527S, 2010AdAst2010E..19S} obtained nightly blocks of three individual exposures in Johnson B and three in Johnson V from May 2016 until January 2018, completed by three exposure in Cousins I since September 2017. The detector is a single 4k$\times$4k back-illuminated thinned CCD with 15\,$\mu$m pixels, providing a field of view of $22 \times 22$ arcmin. The telescope was slightly defocused to minimise scintillation noise and improve the quality of the photometry \citep{Mallonn2016}. The data reduction employed the same ESO-MIDAS routines used for previous monitoring programmes of exoplanet host stars with STELLA/WiFSIP \citep{Mallonn2015,Mallonn_2018}. Bias and flat-field correction was done by the STELLA pipeline. We used SExtractor \citep{1996A&AS..117..393B} for aperture photometry and extracted the flux in elliptical apertures (SExtractor aperture option MAG\_AUTO). Data points of low quality due to non-photometric conditions were discarded. Weighted nightly averages were considered for each filter, resulting finally in a total of 112 measurements in the $B$ filter, 110 in $V$, and 16 in $I$. The use of nightly averages is justified because the noise floor for each night is typically limited by systematic errors in the measurement procedure and calibration of ground-based photometric observations. We avoid giving excessive weight to nights with larger number of measurements by computing the nightly averages and adding a jitter term to account for random errors not included in the model of the observations such as night-to-night calibration errors. This strategy is appropriate because we are interested in rotational modulation timescales, with the few-hour time domain being irrelevant.

The TJO telescope provided photometric data using its main imager MEIA2. The instrument has a single 2k$\times$2k back-illuminated thinned CCD yielding a field of view of $12.3 \times 12.3$ arcmin and a resolution of 0.36 arcsec/pixel. The images were calibrated with darks, bias and flat fields using the ICAT pipeline \citep{2006IAUSS...6E..11C} and differential photometry was extracted using AstroImageJ \citep{AstroImageJ}. A total of 66 and 77 epochs, that is, weighted nightly averages, were obtained with the Johnson-Cousins $B$ and $R$ filters, respectively.

\subsection{Photometric inverse problem} 
\index{inver52} 
\label{inver52}

\subsubsection{Fixed \textsf{StarSim} parameters}

The basic stellar parameters (relevant to select the appropriate atmosphere models) were adopted to be $T_{\rm eff}=5000$\,K and $\log{g}=4.5$, which are close to the literature values (see Table\,\ref{tab:WASPparams}). The stellar inclination was chosen to be $90^{\circ}$, which is a reasonable assumption given the measured values of the planet's orbital inclination and spin-orbit angle. The lifetime of the active elements was assumed to be Gaussian-distributed around $250\pm100$\,days. The choice of this value is done considering the full timespan of the series and the fact that there are two separate epochs covered by our datasets, each of $\sim$200 days in duration. This allows for the presence of different active groups in each epoch, as well as some common regions in both, creating a flexible evolving map of the active regions. Our selection also matches the range of spot lifetimes from 70 to 350\,days described for WASP-52 by \citet{Mancini2017}. Several tests showed that those values are not critical in terms of fitting quality and do not produce any bias in other parameters. Furthermore, we did not consider fitting for differential rotation (which \textsf{StarSim} could) as this would add even more complexity to the parameter space. If significant differential rotation was present, this would be naturally accounted for by the fitting algorithm through the appearance and disappearance of active regions at slowly varying longitudes. The temperature contrast between faculae and the photosphere was fixed to $\Delta T_{\rm{fc}}=50$~K. This is consistent with the results from \cite{1993SSRv...63....1S} \citep[see][for further discussion]{Herrero2016}. As is shown in Sect.~\ref{inv1}, our inversion process yields solutions with insignificant facular coverage independently of the temperature contrast $\Delta T_{\rm{fc}}$. Finally, we set a restriction to the active element colatitudes, namely that we only allowed their presence in one of the stellar hemispheres. This is possible in our case because of latitudinal symmetry. By doing so, we guarantee the absence of spot crossing events when we study planetary transits later in our analysis.

As discussed in Sect.\,\ref{sec:gen_spot_maps}, there are some other parameters that need to be set beforehand. One is the number of iterations of the MCSA algorithm. Ideally, this should be sufficiently large to ensure consistency in the resulting surface map but at the same time be computationally affordable. The number of iterations is also linked to the quantity of active elements considered on the stellar surface, since each element contributes 5 potential parameters. This number has to be sufficient large to generate the inhomogeneities of the stellar surface causing the photometric variability, yet not as large as to overfit instrumental noise. It partially depends on the timespan of the dataset and the lifetime of spots, and the maximum radius of the active elements, which we set to 10 degrees to be consistent with the small spot approximation.

We ran a battery of tests considering different number of iterations and surface active elements. An initial exploration of the parameter space was used to obtain valid guesses of the optimal parameters $P_{\rm rot}$, $\Delta T_{\rm sp}$, $Q$, and $s$ (jitter). We systematically explored combinations of number of iterations, from 500 to 10\,000, and number of spots from 60 to 150 and evaluated the resulting $\ln \mathcal{L}$ values. For each of the explored pairs, we ran 112 realisations starting with different random spot maps, except for the runs with 10\,000 iterations, for which we considered 56 realisations. 

\begin{figure}
\begin{center}
\includegraphics[width=9cm]{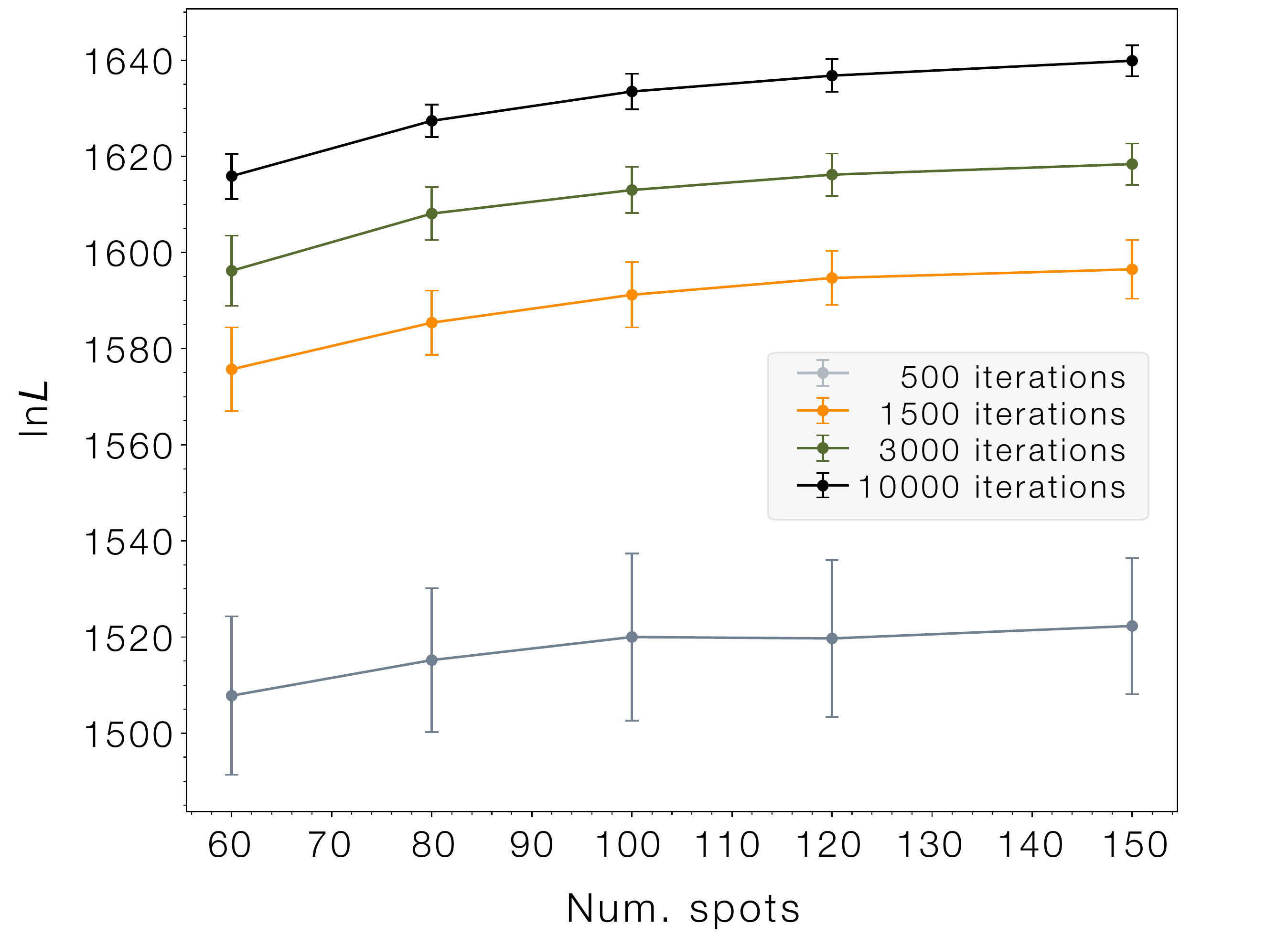}
\end{center}
\caption{Trial tests showing the statistic $\ln \mathcal{L}$ as a function of the number of spots and for several values of the number of iterations of the MCSA algorithm. The size of the error bars corresponds to the 1-sigma intervals around the optimal solutions found from 112 random initial spot maps, except for the case with 10\,000 iterations, where we employed 56.}        
\label{fig:spot_tests}
\end{figure}

The results of the tests are shown in Fig.\,\ref{fig:spot_tests}. It is not surprising to see that the quality of the fits improves with the number of iterations, as it also does with the number of spots. The criterion to select an adequate number of iterations is mostly related to computational cost. For the present problem, numbers above 3\,000 iterations per MCSA step are prohibitive. Nevertheless, 3\,000 iterations already delivers very consistent solutions from random starting points. The final variance of the statistic $\ln \mathcal{L}$ from the sample converged solutions is about 5, which is sufficient to guarantee stable solutions. For lower numbers of iterations, this number increases to 7 (1500) and 17 (500). We note that for 10\,000 iterations (3$\times$ longer computational time) the final variance is approximately 3. Regarding the number of active elements we see that less than 80 spots is clearly insufficient to fit our data (precision \& time span); and we also see that the improvement in likelihood flattens out quite apparently beyond 120 spots. For simplicity arguments, and factoring in again computational costs, we find that using 100 spots delivers sufficiently reliable fits, and note as well that the average $\ln \mathcal{L}$ values for 150 and 100 spots overlap at the 1-sigma level. Thus, our adopted values regarding the number of MCSA iterations and spots was 3\,000 and 100, respectively, and this produces a optimal solutions with a variance of $\sigma_{\ln \mathcal{L}} = 4.8$.

\subsubsection{Fit results} \index{inv1} \label{inv1}

The available datasets consist of multiband photometry and we are especially interested in exploring the parameters related to the chromatic properties of active regions, which play a major role on the characterisation of the effects of activity on transit spectroscopy measurements studied in Sect.~\ref{S4}. The parameters describing stellar properties must be fitted simultaneously to provide consistent solutions, as explained in Sect.~\ref{subsec:optimal_theta}. In our case, besides the 100-spot map, those parameters are the rotation period, $P_{\rm rot}$, which drives the timescale of the variability, the facula-to-spot area ratio, $Q$, which determines the fractional coverage of bright active elements, the temperature contrast of the spots, $\Delta T_{\rm sp}=T_{\rm ph}-T_{\rm sp}$, and the additional noise term or jitter, $s$.

As discussed in Sect.~\ref{sec:gen_spot_maps} and shown in Fig. \ref{fig:flow_chart}, the inversion procedure with \textsf{StarSim} consists of firstly selecting a parameter set drawn from a prior distribution covering the relevant parameter space, and subsequently optimizing random spot maps for each realisation. The resulting statistic $J$ is then used to evaluate the relative quality of each set of parameters. We have shown before that the intrinsic variance of the statistic for our problem is $\sigma_{\ln \mathcal{L}} = 4.8$, which indicates that differences of such order for different parameter sets are statistically indistinguishable. Therefore, the recipe that we adopt is taking the best (larger $J$) solution from the batch and then considering as statistically equivalent those that are within 3-sigma ($\Delta \ln \mathcal{L}_{\rm lim} = 14.4$). We are aware that this is not completely satisfactory because the interval may include some solutions with midly sub-optimal parameter sets, but we adopt this criterion to ensure a statistically significant collection of good solutions at the expense of some increase in the parameter uncertainties. Increasing the number of MCSA iterations or speeding up convergence could make this criterion more stringent. Admittedly, this is a limitation of the current algorithm imposed by the complexity of the optimisation method that should be improved in subsequent releases of \textsf{StarSim}.   

\begin{figure}
\begin{center}
\includegraphics[width=8.7cm]{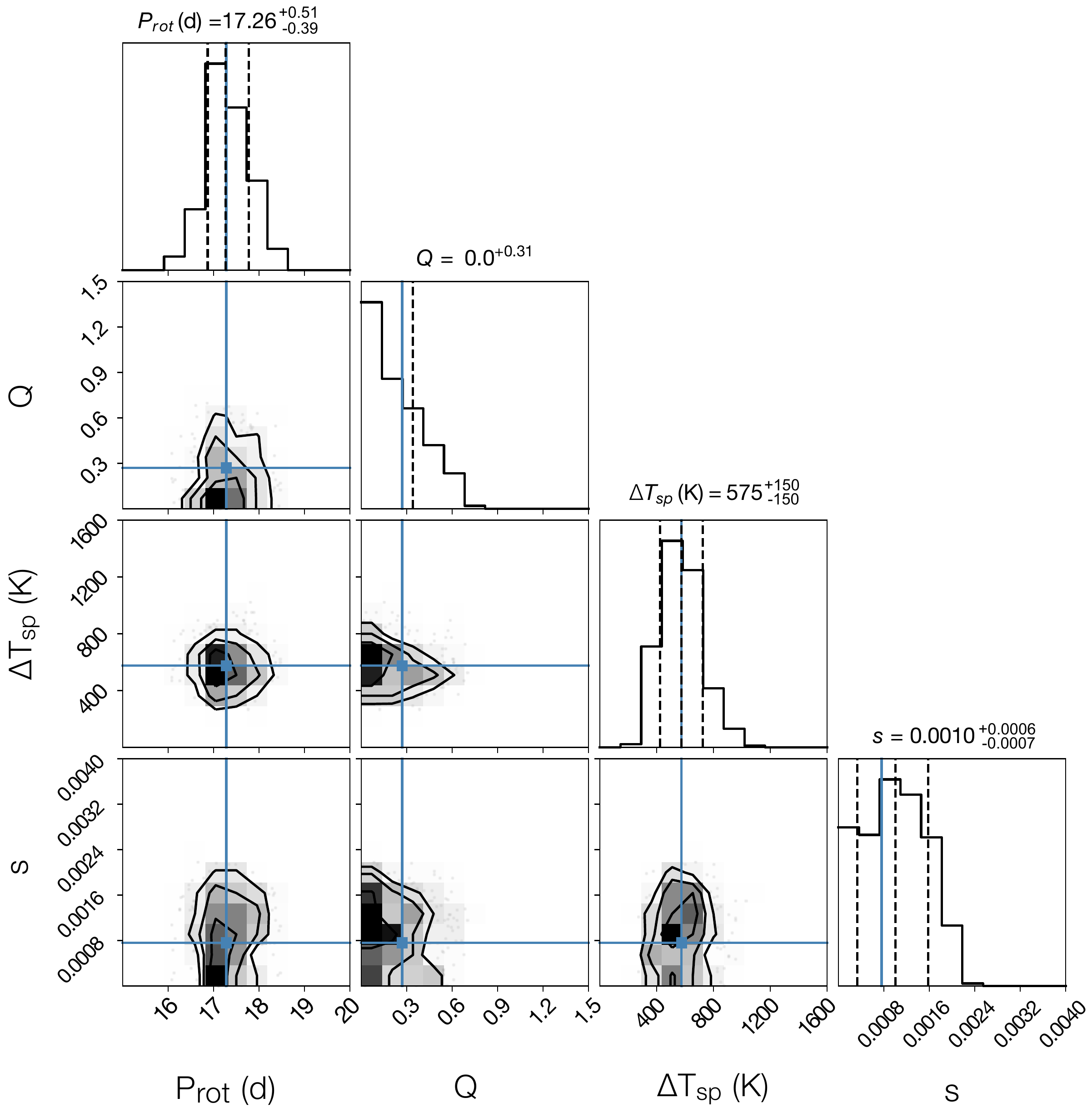}
\end{center}
\caption{Histograms and correlations of 365 statistically-equivalent solutions for the fitted \textsf{StarSim} parameters: the rotational period, $P_{\rm rot}$, the facula-to-spot area ratio, $Q$, the spot-photosphere temperature contrast, $\Delta T_{\rm sp}$, and the photometric jitter, $s$. The dotted lines mark the median and 68\% percentiles of the parameter distributions and the blue lines indicate the parameters of the solution with the best likelihood value found.}
\label{fig:param_fit}
\end{figure}

After initial tests, for the case of WASP-52 we adopted uniform priors on the parameter sets within the ranges given in Table\,\ref{tab:simparams_all2}. The priors are quite narrow to avoid unnecessarily exploring irrelevant parameter space but, as seen below, the intervals are sampling regions beyond 3-sigma from the optimal parameters. A total of 21\,296 realisations were performed and 108 satisfied the likelihood criterion $\Delta \ln \mathcal{L} < 14.4$ from the highest likelihood one. In a second step, we further narrowed down the parameter priors to include only the ranges defined by the 108 good solutions. An additional 2576 realisations yielded another 257 good solutions, leading to a grand total of 365. These are the $\theta$-configurations that were used to build the parameter distributions. The distributions are plotted in Fig.\,\ref{fig:param_fit}, and the relevant optimal parameters and their uncertainties are listed in Table \ref{tab:simparams_all2}, compared with other values from the literature. 

\begin{table}
\centering
\caption{Priors and results from the \textsf{StarSim} inversion of WASP-52 photometry. The values given correspond to the optimal solution and uncertainties are estimated from a sample of 365 equivalent solutions. Literature values for the same parameters are provided when available.}
\label{tab:simparams_all2}
\begin{tabular}{l c c c}
\hline \hline \noalign{\smallskip}
  Parameter & Prior & Value & Ref.\\ 
\noalign{\smallskip} \hline \noalign{\smallskip}
  $P_{\mathrm{rot}}$ (days) & $\mathcal{U}(16.0,18.5)$ & $17.26^{+0.51}_{-0.39}$ & This work \\
  \noalign{\smallskip}
     & & 11.8$\pm$3.3$^{(a)}$ & He13 \\  
     & & 16.40$\pm$0.04 & He13 \\
     & & 15.53$\pm$1.96 & Ma17 \\
     &  & 17.79$\pm$0.05 & Lo17 \\
     &  & 18.06$\pm$0.2  & Br19 \\ 
  \noalign{\smallskip} 
  $\Delta T_{\rm sp}$ (K) & $\mathcal{U}(50,2000)$ & 575$\pm$150 & This work  \\
  \noalign{\smallskip} 
  & & 1250--1500 & Ki16 \\ 
    & &  $\sim$ 270 & Ma17 \\
    & & $\sim$ 950 & Br18 \\
    & &  250 & Al18 \\
    & &  2\,230 & Br19 \\     
  \noalign{\smallskip} 
  $Q^{(b)}$ & $\mathcal{U}(0,3)$ & $<0.31$ (1-$\sigma$) & This work \\
  \noalign\\
  $s$ & $\mathcal{U}(0,0.0045)$  & $0.0010^{+0.0006}_{-0.0007}$ & This work \\
  \noalign{\smallskip} \hline \hline
\end{tabular}
\tablefoot{
(a) Obtained from spectral line broadening;
(b) $Q$ is unilaterally distributed.\\ 
{\bf References:} 
He13: \citet{2013A&A...549A.134H}; 
Ki16: \citet{Kirk_2016}; 
Ma17: \citet{Mancini2017}; 
Lo17: \citet{Louden_2017}; 
Al18: \citet{Alam_2018}; 
Br18: \citet{Bruno_2018}; 
Br19: \citet{Bruno_2019}. \\
}
\end{table}

Our model favours a heterogeneous surface dominated by dark spots with a temperature contrast of $575\pm150$\,K with respect to the surrounding photosphere. This value is intermediate to those found in the literature, which can separated into two groups, one with low temperature contrasts of $\sim 250$\,K and another one with contrasts $\gtrsim 1000$\,K. The temperature contrast we find is generally lower than found by \citet{Andersen2015} for K dwarfs and also lower than the prediction by the relationship of \cite{Berdyugina2005}, which would give $\sim 1300$\,K for a K2\,V star. This is, however, not surprising since the strong degeneracy between spot temperature contrast and filling factor can severely bias some determinations. We believe that our value, based on multi-colour photometry, provides a robust estimate. As already mentioned, the results from our analysis do not support a significant presence of bright faculae as expected for rapidly-rotating young K dwarfs \citep{Radick1983,Lockwood2007}. The 365 best solutions indicate a unilateral distribution consistent with $Q=0$. We furthermore find a rotation period of $17.26^{+0.51}_{-0.39}$ days, which is in marginal agreement with the values found by \citet{Louden_2017} and \citet{Bruno_2019}.

\begin{figure*}
\begin{center}
\includegraphics[width=17.0cm]{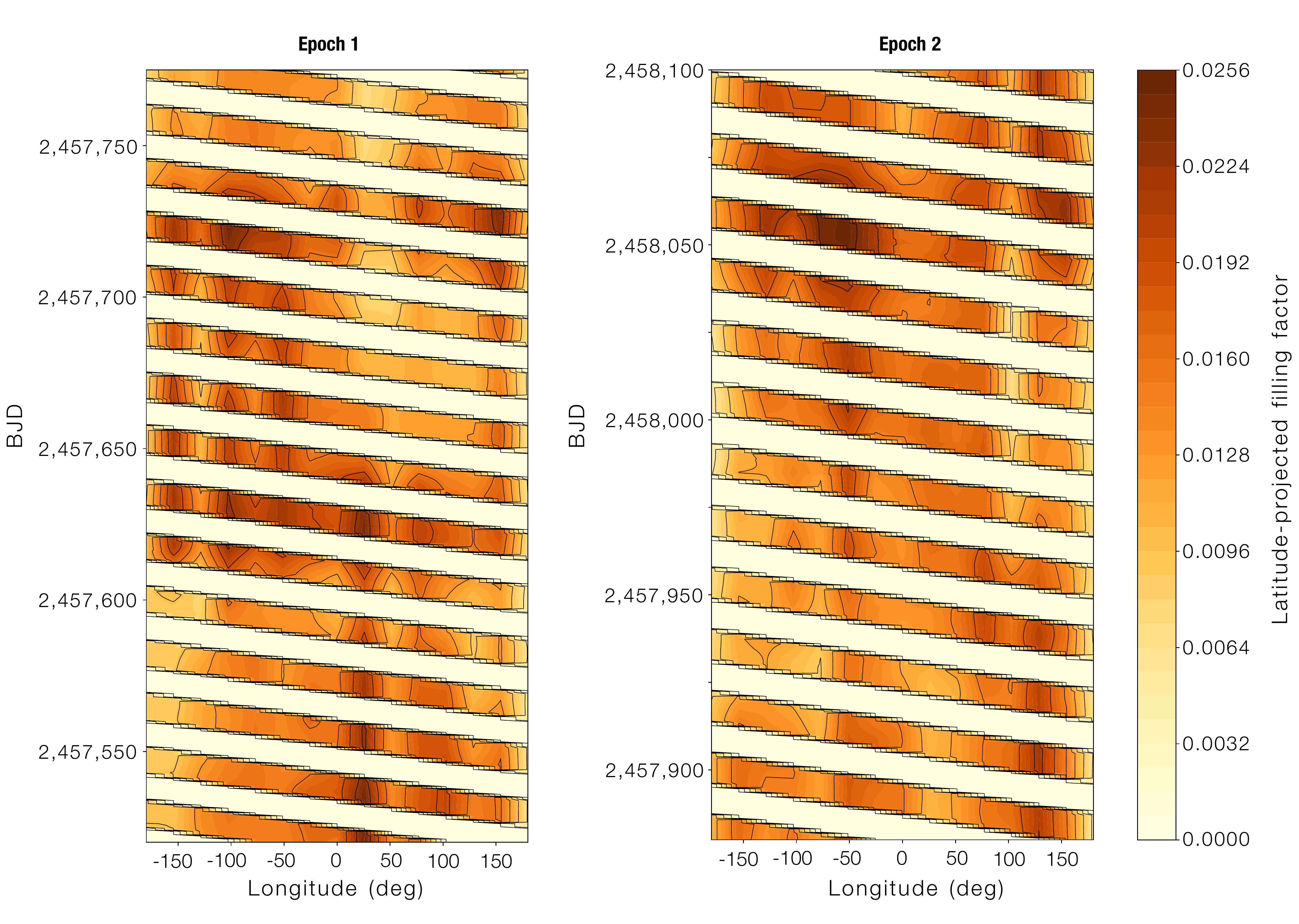}
\end{center}
\caption{Longitudinal spot filling factor projected on the stellar equator of WASP-52 as a function of time, covering photometric epochs in 2016 and 2017 and corresponding to the average of 21 optimal solutions. Only stellar longitudes visible at each time are shown (hence the band structure). The colour scale indicates the fractional latitude-projected spot coverage for each 15-degree longitude bin. 
}
\label{fig:active_longitudes}
\end{figure*}

\subsubsection{The evolving surface of WASP-52}
\label{surface_map}

In addition to the optimal parameters $\hat{\boldsymbol{\theta}}$ and their uncertainties, our inversion realisations also provide a picture of the stellar surface as a function of time. From the 365 accepted solutions we selected only those satisfying $\Delta \ln \mathcal{L} < 4.8$ with respect to the best likelihood value. This yields 21 maps, which should be a good representation of the optimal model considering both the intrinsic randomness of the map inversion from the MSCA algorithm and the statistical variance of the stellar parameters. Fig.\,\ref{fig:active_longitudes} shows the longitudinal spot filling factor projected on the stellar equator, averaged for the 21 optimal maps, as a function of time. We perform a latitudinal projection because of the degeneracy present (see Sect.\,\ref{sec:optimization}). The longitudes in the map are plotted in the reference frame of the rotational period found in our analysis (see Table~\ref{tab:simparams_all2}). The map suggests that there is a clear dominant active region at a longitude of $\sim$30 deg at the beginning of the first epoch, which changes into a more complex longitudinal pattern later in the season, and finally the spottedness level suffers a suddenly decrease. Active regions typically last for $\sim$8 rotations or $\sim$140 days. The second epoch appears to show a better-defined dominance of spot regions, with two alternating active longitudes at about $\sim$130 deg and $\sim-$50 deg, that is, in almost perfect opposition. The distribution does not indicate signs of differential rotation along the sampled period of time. The existence of measurable differential rotation would be seen in Fig.\,\ref{fig:active_longitudes} through the presence of two or more active regions with drifting longitudes. We do not see such effect in the time-series of WASP-52 thus probably indicating that either differential rotation is negligible or that dominant active regions appear at similar latitudes. Such dominant regions are used by the fitting algorithm to define the rotation period.

\begin{figure*}
\begin{center}
\includegraphics[width=18.0cm]{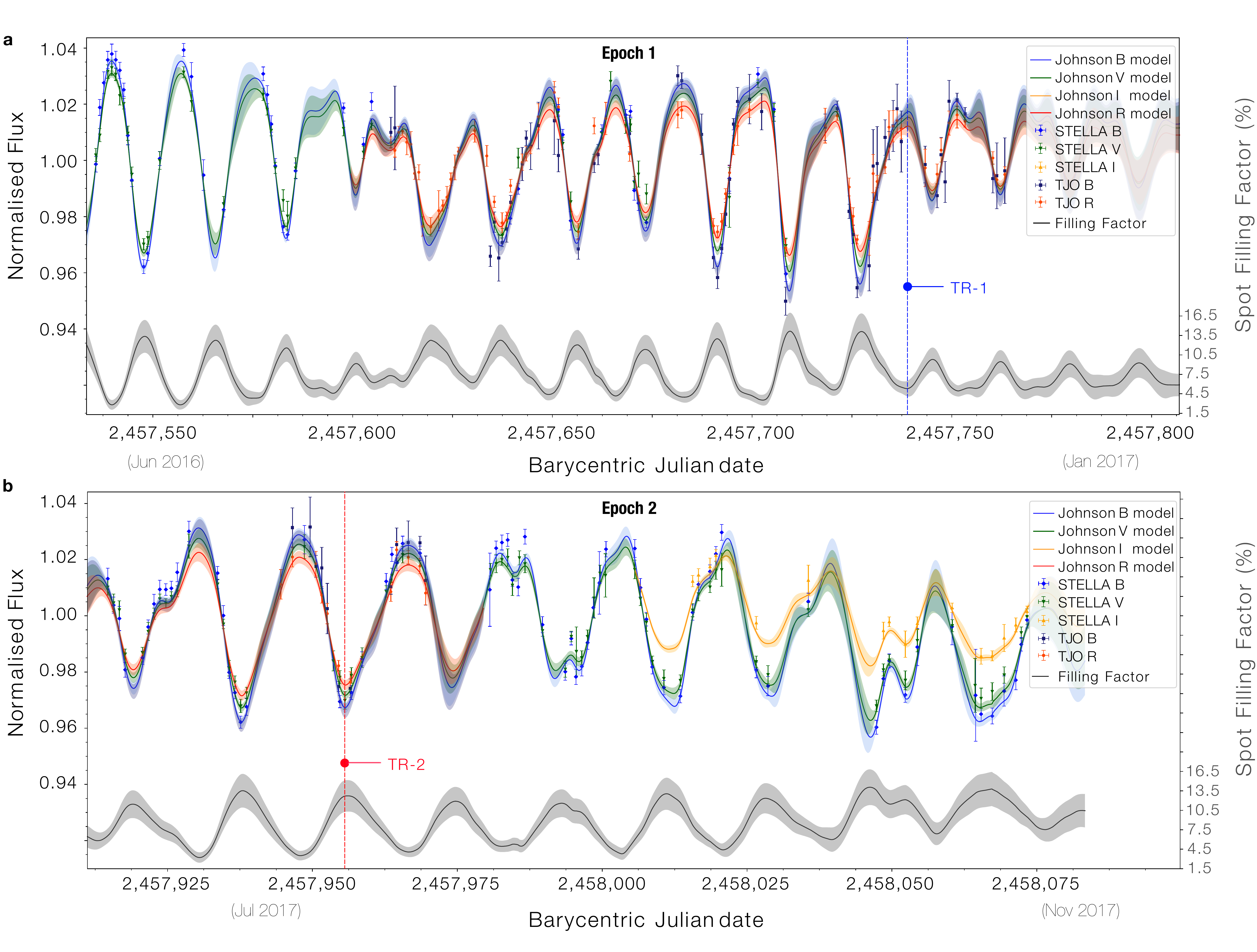}
\end{center}
\caption{\textsf{StarSim} model fits to multiband photometric ground-based observational data as described. Solid curves represent the mean of $21$ optimal solutions of the inverse problem. The shaded bands indicate the 1-$\sigma$ ranges. The gray line at the bottom is the projected spot filling factor of the maps, also showing the mean and 1-$\sigma$ ranges. TR1 and TR2 indicate transit events as discussed in Sect.\,\ref{S4}.}
\label{fig:lc_fits}
\end{figure*}

The best fits to the observational multiband photometry using the modelled parameters of the star and their associated spot maps are shown in Fig. \ref{fig:lc_fits}, together with the multiband photometric measurements. The solid lines show the average models resulting from the optimal 21 surface maps, and the shaded bands show the 1-$\sigma$ range. The two epochs as shown in the top and bottom panel are separated by a $\sim$ 140-day gap. The grey curves at the bottom of each panel show the projected spot filling factor, also showing the average model and the 1-$\sigma$ range. During our observational campaign, total filling factor values covered a range $\sim 3$\%--$14$\% along an observational timespan of $\sim 600$ days in 2016--2017. This implies flux modulations of $\sim 4$\%--$7$\% due to time-varying spot coverage.

\section{Chromatic spot effects on simulated transits} \index{S4} \label{S4}

\subsection{Transit depth variability due to spots}

For magnetically-active stars showing spots on their surface, the transit depth caused by a planet passing in front of the host star depends both on the planet effective radius and the possible inhomogeneities on the stellar photosphere. Using our nomenclature in Sect.\,\ref{foundations}, it is straightforward to show that a planet producing a depth $\mathcal{D}_0$ when transiting an immaculate star, produces a transit depth $\mathcal{D}_{\rm sp}$ when crossing a spotted star, where
\begin{equation}
\label{eq:rackham}
\mathcal{D_{\rm sp}}({\lambda}) =\frac{\mathcal{D}_{0}}{1 - \delta_{\mathrm{ sp}}\Phi^{-1}(\mathcal{B},\Delta T_{\rm sp})},
\end{equation}
with $\delta_{\rm sp}$ being the total spot filling factor. Such very simple model illustrates the chromaticity of the problem, since the function $\Phi$ is wavelength dependent. This has been identified as a source of systematic effects for transmission spectroscopy. For this reason, studying the imprint of active regions is of crucial importance for future instruments and projects aiming at the study of exoatmospheres such as, for example, the JWST and Ariel space missions. 

We can use \textsf{StarSim} to estimate the impact of spots on the transit depth of WASP-52, by simulating transits of the planet. According to the ephemeris of WASP-52, a total of 112 and 104 transits occurred during the time span covered by the first and second seasons of our light curves, respectively. We used the \textsf{StarSim} model of stellar activity obtained in Sect.\,\ref{inver52} for WASP-52. It is important to emphasise that we solely focus on the effects of unocculted star spots, whose properties cannot be constrained from the transit itself. Therefore, in our simulations we avoided planet-spot crossings by positioning active regions outside the swath covered by the planet during the transit. Occulted spots during transits may also have a significant impact on the planet properties. However, if photometry of sufficient precision and time cadence is available (such as expected for JWST and Ariel), spot-crossing events could be identified from the transit observations themselves and potentially modelled. Unocculted spots, on the contrary, leave no visible signature. 

We define the transit depth as the difference between the out-of-transit and the in-transit (at transit mid-time) observed flux. The out-of-transit is the flux of the model without transit calculated at the mid-transit time. Hence, both stellar spots and limb darkening effects cause variations that depend on wavelength. We take this approach for computational simplicity, because we only need to evaluate the flux at the transit mid point and not the full transit event. The transit depth can then be written as
\begin{equation}
\mathcal{D'}({\lambda}) = \mathcal{D}_0 + \mathrm{LD}(\lambda) + \mathrm{SP}(\lambda),
\label{eq:chromatic_effects_1} 
\end{equation}
\noindent
where, $\mathrm{SP}(\lambda)$ and $\mathrm{LD}(\lambda)$ are the additive chromatic effects induced by spots and by stellar limb darkening, respectively, and $\mathcal{D}_0$ is the non-chromatic (asymptotic) transit depth, which we assume to be constant, that is, we do not consider here the planet atmosphere. This can be written as
\begin{equation}
\mathcal{D}_0=\left(\frac{r}{R_*}\right)^2, 
\end{equation}
\noindent
where $r$ and $R_*$ are the planetary and stellar radii, respectively. In this paper, we assume $\mathcal{D}_0=0.0253$ for WASP-52~b \citep{Ozturk_2019}. 

The transit depth including chromaticity effects can then be expressed as
\begin{equation}
\mathcal{D'}({\lambda}) =\left(\frac{r+\Delta r}{R_*}\right)^2,
\end{equation}
where $\Delta r$ accounts for the effect of spots and limb darkening on the planetary radius as a function of wavelength. Expanding Eq.\,(\ref{eq:chromatic_effects_1}), and neglecting second order terms, we derive
\begin{equation}
\frac{\Delta r}{R_*} = 
\frac{1}{2}\,\frac{1}{\sqrt{\mathcal{D}_0}}\,\big[\mathrm{LD}(\lambda) + \mathrm{SP}(\lambda) \big].
\label{eq:total_chromatic_effect} 
\end{equation}
\noindent
Therefore, we can separate the chromatic contributions due to limb darkening ($\Delta r_{\rm LD}$) and spots ($\Delta r_{\rm SP}$) as,
\begin{equation}
\frac{\Delta r_{\mathrm{LD}}}{R_*} =  \frac{1}{2}\,\frac{1}{\sqrt{\mathcal{D}_0}}\,\mathrm{LD}(\lambda), \, 
\label{eq:chromatic_effects_LD} 
\end{equation}
\noindent
and
\begin{equation}
\frac{\Delta r_{\mathrm{SP}}}{R_*} =  \frac{1}{2}\,\frac{1}{\sqrt{\mathcal{D}_0}}\,\mathrm{SP}(\lambda).
\label{eq:chromatic_effects_SP} 
\end{equation}
\noindent

\begin{figure*}
\begin{center}
\includegraphics[width=15.5cm]{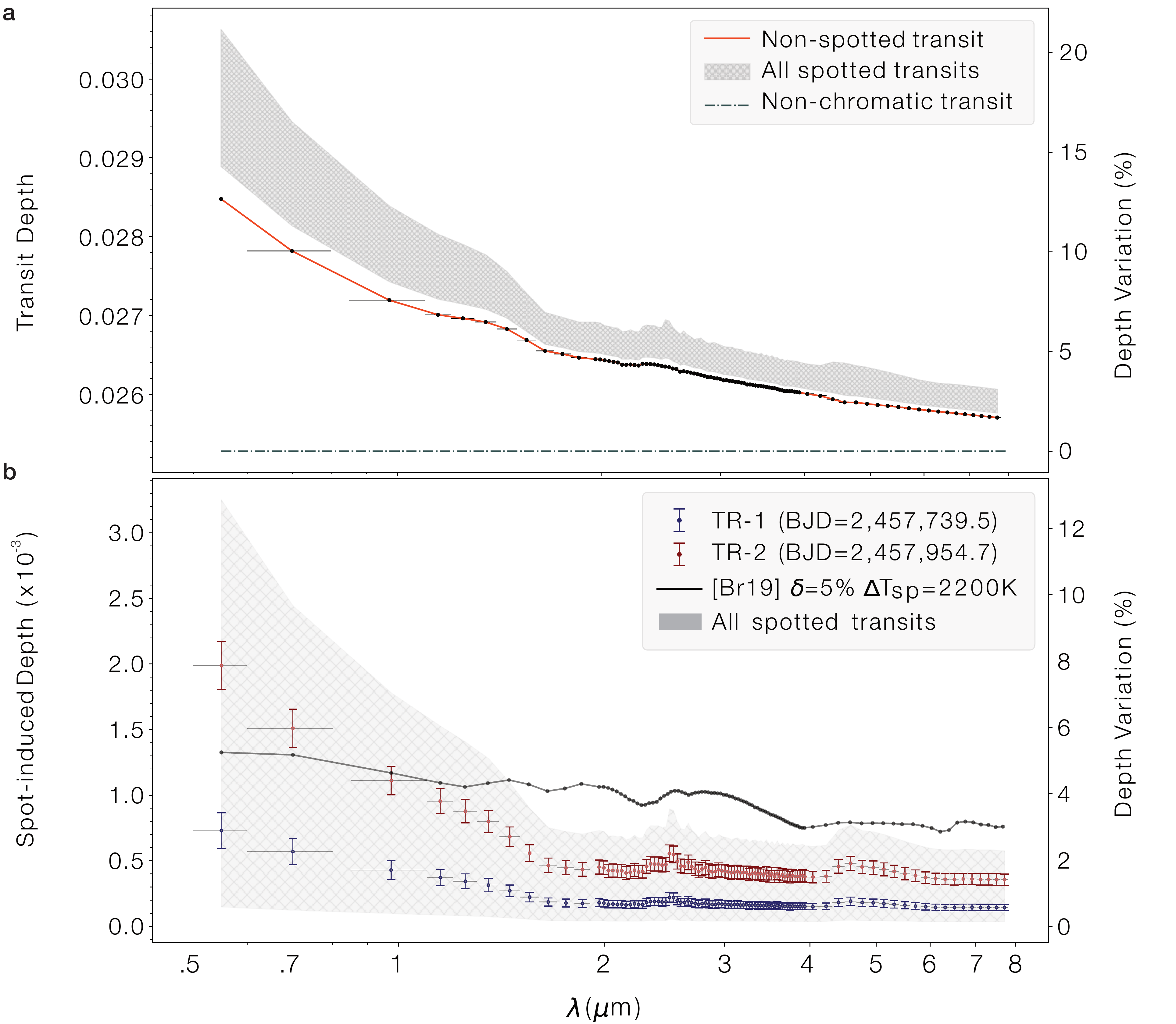}
\end{center}
\caption{\textbf{(a)} Transit depth as a function of wavelength for the WASP-52 system simulated assuming an immaculate surface (red) and including spots in the photosphere (gray band), avoiding transit spot crossings. The gray area illustrates the total coverage from all simulated transits for the 21 optimal spot configurations, clearly showing the transit depth chromatic bias produced by non-occulted spots. The dotted blue line is the non-chromatic transit depth assuming a uniform photosphere ($(r/R_*)^2$) according to the parameters in \citet{Ozturk_2019}. \textbf{(b)} Chromatic signature on transit depths due to spots after subtracting limb-darkening effect (in red in top panel). TR-1 and TR-2 show two example transits corresponding to low- and high-activity -- marked with vertical lines in Fig. \ref{fig:lc_fits} -- respectively, and the gray band represents the area covered by all simulated transits. The black solid line with symbols shows the chromatic effect estimated using stellar spot parameters derived by \cite{Bruno_2019}.}
\label{fig:depth_curves}
\end{figure*}

Figure\,\ref{fig:depth_curves} shows the results of the simulations. The shaded gray region in panel (a) displays the range of transit depth values as a function of wavelength for the simulated transits comprised within the timespan of our datasets. Such transits differ in the amount and distribution of spots in the projected stellar surface. As a reference, the red line shows the transit depth in the case of an immaculate star. In this latter case, $\mathcal{D}_{imm}'=\mathcal{D}_0+\mathrm{LD(}\lambda\mathrm{)}$, the wavelength dependence is solely due to the limb darkening effect, which for WASP-52 produces transit depths from 0.0285 to 0.0256 in flux ratio units from the visual to the infrared bands, respectively. As reference, the dot-dashed blue line indicates the expected transit depth for a uniform planetary disc corresponding to $\mathcal{D}_0=(r/R_*)^{2}=0.0253$. The shaded grey region in panel (b) shows the same results when removing the contribution of limb darkening, $\Delta \mathcal{D}' = \mathrm{SP(}\lambda\mathrm{)} = \mathcal{D}_{sp}' - \mathcal{D}_{imm}'$, therefore only including the chromatic effect due to the presence of spots. The results indicate that transit depth variations of WASP-52\,b due to spots vary from $\sim$2$\times10^{-4}$ to $\sim$3$\times10^{-3}$ in the visual region, and from $\sim$10$^{-4}$ to $\sim$7$\times10^{-4}$ in the NIR region. The range in the variations is caused by both different spot filling factors during each transit, and the uncertainty on the exact distribution of spots. In other words, our results show that the transit depth of WASP-52~b can be overestimated by up to $\sim$12\% and $\sim$3\% in the visual and NIR bands, respectively, thus potentially affecting the retrieval of planetary atmosphere parameters, which could be one or two orders of magnitude smaller \citep{Kreidberg_2018}.

To understand the impact of different spot configurations we selected two transit events representing low and high-activity levels at well-sampled epochs of the light curves. These are labelled in Fig.\,\ref{fig:lc_fits} as TR-1 at BJD=2,457,739.475 (16 Dec 2016, low-activity case), when the projected filling factor of spots is estimated to be close to its minimum value around 5\%, and as TR-2 at BJD=2,457,954.698 (20 Jul 2017, high-activity case), when spot projected filling factor reaches $\sim$12\%. Panel (b) of Fig. \ref{fig:depth_curves} shows the transit depth variation of TR-1 and TR-2 in blue and red, respectively, for the 21 optimal maps, with vertical error bars indicating the 1-$\sigma$ variance. These test cases clearly show that the strength of the chromatic effect is $\sim$3 times larger when the projected spot filling factor is larger by about the same factor. This implies that the effect is also correspondingly larger for stars showing higher levels of stellar activity. 

Table \ref{tab:summary_table} provides the statistics of the solutions for the TR-1 and TR-2 transit cases for different wavelengths, encompassing the visual channels of the Ariel mission ($\sim$550\,nm and $\sim$705\,nm), the infrared ($\sim$975\,nm) and the low resolution spectrographs (NIR, AIRS). We computed the mean transit depth variation induced by spots, $\langle\mathrm{SP}(\lambda)\rangle$, and its corresponding planetary radius variation $\langle\Delta r / R_{*}\rangle$, along with the standard deviation $\sigma$ induced by the uncertainty in the determination of the spot map. The statistics reveal that, even for the relatively low stellar activity level of WASP-52, with a spot filling factor of $\sim$5\%, the signature of stellar activity on the transit depth can be close to $10^{-3}$ of the flux in the visual bands and $\sim$2$\times10^{-4}$ in the NIR bands. This translates into relative planetary radius changes from $\sim$2$\times10^{-3}$ to $\sim$5$\times10^{-4}$, depending on the wavelength considered, which are of the same order as the typical signature of exoplanet atmospheres on transmission spectra. Furthermore, at higher activity level such as that of TR-2, the effect of the starspots is about twice as large. This illustrates the complication of studying atmospheres of exoplanets orbiting relatively active host stars.

\begin{table*}[ht]
\centering
\caption{Mean values of transit depth variations due to spots, $\mathrm{SP}(\lambda)=\Delta\mathcal{D}^{\prime}$, on WASP-52 simulated transits. The corresponding planet relative radius variations, $\frac{\Delta r_{SP}}{R_*}$, are  computed following Eq.\,(\ref{eq:chromatic_effects_SP}). Values for the low-activity transit TR-1 and the high-activity transit TR-2 cases are provided. $\lambda_{\textrm{eff}}$ refers to the central wavelength of the corresponding channel.}
\label{tab:summary_table}

\begin{tabular}{ll|cccc}
\noalign{\smallskip}
\hline
\hline
\noalign{\smallskip}
\multicolumn{1}{c}{\multirow{2}{*}{Instrument}} & \multicolumn{1}{c|}{\multirow{2}{*}{$\lambda_{\textrm{eff}}$ ($\mu$m)}} & \multicolumn{2}{c}{$ (\langle \mathrm{\mathrm{SP}}(\lambda)\rangle \pm \sigma$) $\times 10^{-3} $} & \multicolumn{2}{c}{$ (\langle\frac{\Delta r_{\mathrm{SP}}}{R_*}\rangle\pm\sigma)$  $\times 10^{-3} $} \\

\multicolumn{1}{c}{} & \multicolumn{1}{c|}{} & \multicolumn{1}{c}{\small{low-activity TR-1}} & \multicolumn{1}{c}{\small{high-activity TR-2}} & \multicolumn{1}{c}{\small{low-activity TR-1}} & \multicolumn{1}{c}{\small{high-activity TR-2}} \\ 
\noalign{\smallskip}
\hline \noalign{\smallskip}
VIS-Phot & $0.55$  & $0.73 \pm 0.13$ & $1.99 \pm 0.18$ & $2.30 \pm 0.41$ & $6.26 \pm 0.57$  \\
FGS-1    & $0.7$   & $0.57 \pm 0.10$ & $1.51 \pm 0.15$ & $1.79 \pm 0.31$ & $4.75 \pm 0.47$ \\
FGS-2    & $0.975$ & $0.43 \pm 0.07$ & $1.11 \pm 0.11$ & $1.35 \pm 0.22$ & $3.49 \pm 0.35$ \\
NIR-Spec & $1.525$ & $0.22 \pm 0.04$ & $0.56 \pm 0.06$ & $0.69 \pm 0.13$ & $1.76 \pm 0.19$ \\
AIRS-Ch0 & $2.95$  & $0.17 \pm 0.03$ & $0.43 \pm 0.05$ & $0.53 \pm 0.10$ & $1.35 \pm 0.16$ \\
AIRS-Ch1 & $5.875$ & $0.15 \pm 0.02$ & $0.37 \pm 0.04$ & $0.47 \pm 0.10$ & $1.16 \pm 0.13$ \\

\noalign{\smallskip}
\hline
\hline
\end{tabular}

\end{table*}

\subsection{Correcting transit depth for stellar activity}
\label{sec:transit_depths}

Various of methodologies to correct the impact of stellar activity on transmission spectroscopy of transiting planets have been suggested \citep[see e.g.][]{Knutson2012,Pont2013,Alam_2018}. Typically, they are based on measuring the photometric variability of the host star, from which the spot filling factor and the temperature contrast or a combination of both are estimated. Furthermore, \cite{2018ApJ...853..122R} claim that the methodologies applied to calculate these corrections could still be biased as they are usually computed assuming a single spot on the surface of the star. The problem is that photometric variations do not generally provide a realistic representation of the spot filling factor throughout the rotation cycle. For example, a uniformly densely spotted star could produce a bias on the transit depth while not producing any observable photometric variability. Using semi-empirical relations between the filling-factor of spots and the photometric variability of stars, \cite{2018ApJ...853..122R} conclude that the chromatic effect on the transit depth due to spots could be several times larger than the signal of atmospheric features, thus greatly difficulting the general use of transmission spectroscopy for stars with some level of magnetic variability. 

We demonstrate here that using multiband photometric light curves contemporaneous to transit observations, we can overcome this problem, at least partly. The advantage of using several photometric bands is that, as discussed in the previous section \citep[see also][]{Mallonn_2018}, it is possible to independently estimate the zero point with respect to the unspotted photosphere ($z$), and thus the absolute spot filling factor (including any persistent level of spottedness during the rotation cycle), and the temperature contrast. This provides extremely valuable information to correct transit depth variations due to spots. We can actually use our \textsf{StarSim} simulations of WASP-52 to investigate at which level of accuracy we can correct the effects of starspots on transit data.

The \textsf{StarSim} simulations of the WASP-52 planetary system discussed in the previous section reveal that, without additional information allowing us to infer the distribution and filling factor of spots during the transit, the planet-to-star radius cannot be determined with an accuracy better than a few percent. This is because in that case, we do not know at which rotation phase of the star the transit is observed, and therefore, we can only estimate a range of possible filling factors based on the amplitude of the photometric modulation. However, if we know the rotation phase at the time of transit, and we can estimate the spot properties and distribution from light curves, we can infer the correction that needs to be applied to the relative radius of the exoplanet (or transit depth) with an accuracy of order $10^{-4}$, as shown in Table\,\ref{tab:summary_table}.

\begin{figure*}
\begin{center}
\includegraphics[width=18cm]{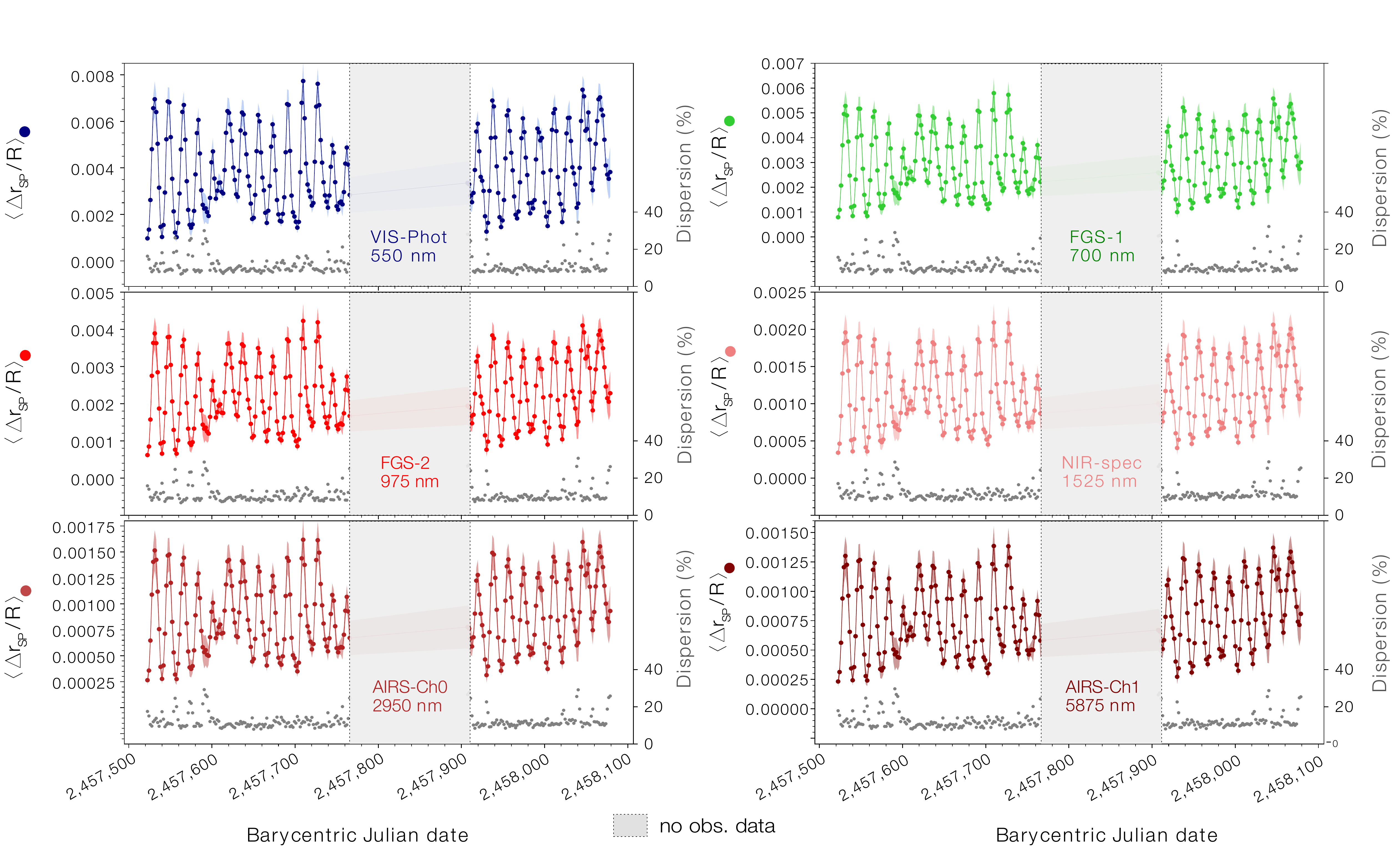}
\end{center}
\caption{Effect of unocculted dark spots on the planetary radius $\Delta r_{SP}/R_*$ over time for different wavelength bands of the Ariel mission. The solid lines correspond to the average value of 21 optimal fits, and the 1-$\sigma$ confidence level is shown as a light-shaded band. Solid dots indicate the timestamps of WASP-52~b transits according to the available ephemeris. Note the different vertical scale for each wavelength panel. Gray dots at the bottom of the panels represent the variance ($\sigma$, as a percentage value) of the results calculated from the different maps, as labelled in the right axis. The gap in the middle of each plot corresponds to the period between observational seasons without available photometric data (see Fig\,\ref{fig:lc_fits}). }
\label{fig:chromatic_effects}
\end{figure*}

Figure \ref{fig:chromatic_effects} illustrates the sequence of spot corrections on the transit depth as a function of time. We plot the planet radius variation due to unnoculted spots, $\Delta r_{\mathrm{SP}}/R_*$, following our \textsf{StarSim} model describing the photometric light curve in Fig.\,\ref{fig:lc_fits}. Each panel corresponds to a different wavelength band of interest for the Ariel mission, as an example. To test all possible filling factors, we consider transits occurring at any time, but actual WASP-52~b transits are indicated as filled dots. The average value of all the simulations performed is displayed as a solid line. It is clear from this figure that the variation of the apparent planet radius depends on the projected spot filling factor and follows the stellar rotation period. The relative effect on the radius diminishes towards longer wavelengths due to the lower flux contrast of spots, from an average of $\sim$4$\times10^{-3}$ to $\sim$8$\times10^{-4}$ at 550\,nm (VIS-Phot) and 6\,$\mu$m (AIRS-Ch1), respectively. In all cases it is possible to estimate the unocculted spot correction to the planetary relative radius with a precision close to 10\%, estimated as the standard deviation ($\sigma$) of the sample of 21 optimal fits to the light curves. As explained above, this residual uncertainty is due to the variance of the stellar parameters and the randomness of the determination of the active region map.

\begin{figure}
\begin{center}
\includegraphics[width=9cm]{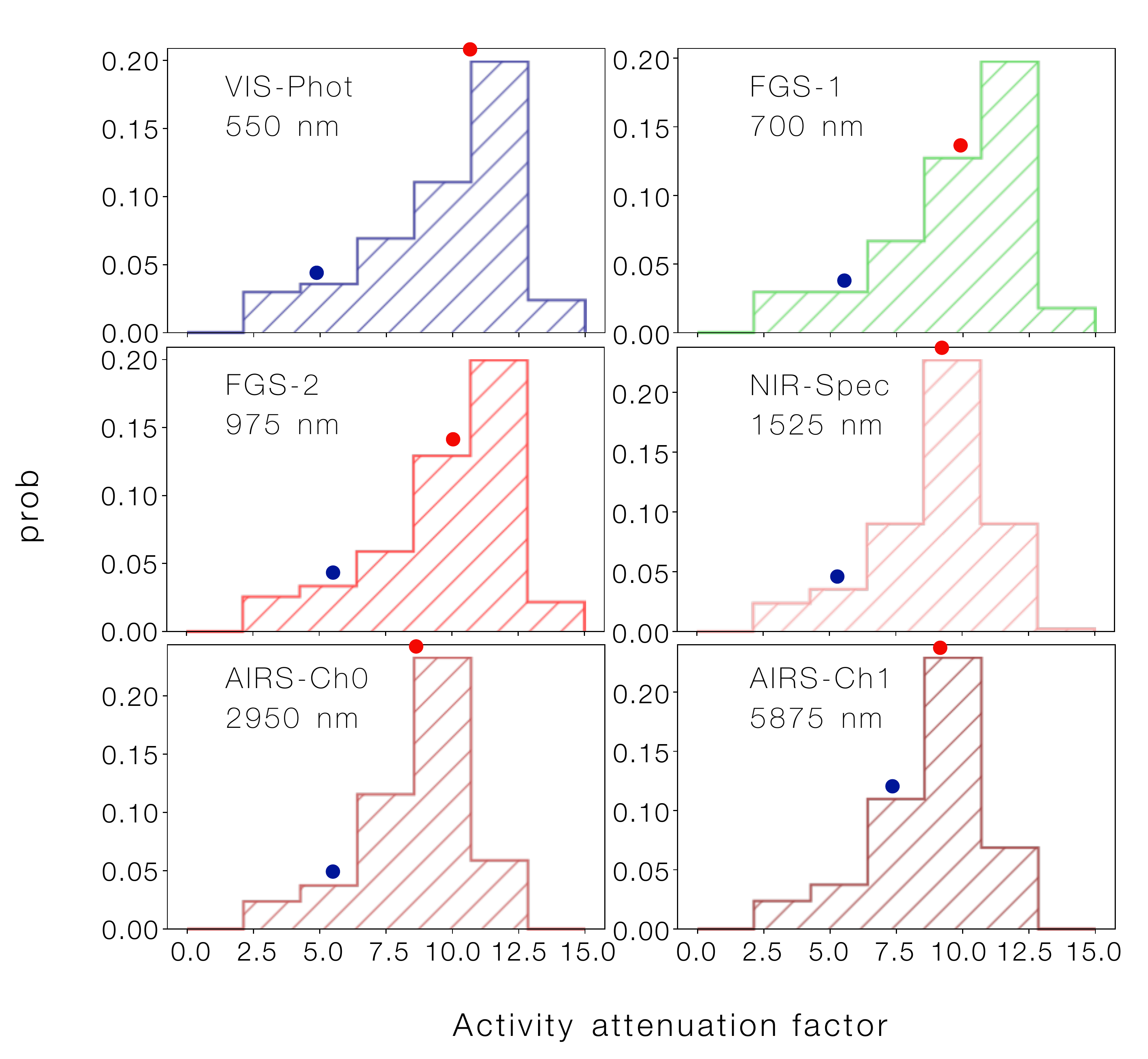}
\end{center}
\caption{Distribution of activity attenuation factors from \textsf{StarSim} modelling considering all transit simulations. This factor is calculated as the ratio between the mean effect of spots on the transit depth and the uncertainty of the correction given by \textsf{StarSim} model ($\langle$SP($\lambda$)$\rangle$/$\sigma_{\rm SP}$). Blue and red dots indicate the position in the histograms of the example transits TR-1 and TR-2, respectively.}
\label{fig:histograms}
\end{figure}

It is interesting to quantify the improvement on a realistic transit depth determination that we can reach using this approach. We measure this improvement through what we call activity attenuation factor, which we compute as the ratio between the effect of spots on the transit depth and the uncertainty of the correction resulting from the \textsf{StarSim} model ($\langle$ SP($\lambda$)$\rangle$/$\sigma_{\rm SP}$). The distribution of attenuation factors for the different simulated transits is shown in Fig.\,\ref{fig:histograms}, and this is a measure of the ability of the \textsf{StarSim} model to correct starspot effects. The plots indicate that, using simultaneous photometry, we can correct the effects of spots on the relative radius of WASP-52~b by factors between 2 and 15 depending on the spot filling factor and the accuracy and coverage of the contemporaneous photometric monitoring. Lower attenuation factors mainly correspond to transits occurring when the projected filling factor of spots is smaller and having poor photometric coverage. On the other hand, larger factors are mainly due to transits with greater spottedness at times of well-sampled photometric light curves. On average, we can expect an attenuation factor around 10 on the relative radius measured for WASP-52~b.

The simulations of this planetary system show that, using contemporaneous photometric data, we are able to estimate transit depth corrections due to unnoculted spots with uncertainties of around few times $10^{-5}$ in the NIR, regardless of the fraction of photosphere covered by spots (see error bars in the first two columns of Table\,\ref{tab:summary_table}). And we recall that this corresponds to a star displaying a modulation of about 7\% in flux in the visible band, and a spot filling factor of about 3--14\%. 

\subsection{Comparison with alternative activity correction methods for WASP-52~b}
\label{sec:other}

Several approaches have been presented to account for and correct out the chromatic effects of activity in transit observations. Instead of fitting photometric light curves, \cite{2019AJ....157...96R} estimate the filling factor of FGK spectral type stars from the peak-to-valley variability measured from \emph{Kepler} photometry using the simple analytical form in Eq.\,(\ref{eq:rackham}). The temperature contrast of the active regions and the projected filling factor of spots are strongly degenerate when only a single passband is used to measure photometric variability. To overcome this problem, \cite{2019AJ....157...96R} assume a temperature for the spots consistent with empirical values reported by \cite{Berdyugina2005}. The authors also assume a uniform distribution of dark spots over the stellar photosphere. If we apply this approach to WASP-52, the temperature contrast of spots would be 1290~K. However, in spite of being significantly larger than the value we find in our fits, with this value it is not possible to reproduce the $\sim$7\% peak-to-peak photometric variability of our light curves \citep[see Fig.\,2 in][]{2019AJ....157...96R} when assuming uniform spot coverage. Unphysically high spot coverage levels, a much larger spot temperature contrast or non-uniform distributions would be needed to explain the large observed amplitude. In addition to the different temperature contrast we obtain, the surface distribution of active regions derived with \textsf{StarSim} is distinctly non-uniform (see Fig.\,\ref{fig:active_longitudes}).

\cite{Bruno_2019} also estimate the properties of starspots for WASP-52, following the procedure of \cite{Huitson2013} and the light curve normalisation factor suggested by \cite{Aigrain2012}. Based on data in \cite{Alam_2018}, the authors fit both parameters from ASAS-SN and AIT photometry and transit spectroscopy from HST/STIS, WFC3 IR, and Spitzer/IRAC. They obtain temperatures $<$3000~K for the starspots and a $\sim$5\% filling factor. Although the filling factor is consistent with our results, the temperature of spots is much larger than the value we measure from multiband photometry. 
The chromatic signature produced by the parameters of \cite{Bruno_2019} is plotted in panel (b) of Fig.\,\ref{fig:depth_curves} (black solid line) using Eq.\,(\ref{eq:rackham}). At wavelengths below $\sim$1.5\,$\mu$m), the resulting values are within the region allowed by our \textsf{StarSim} runs, and below 1\,$\mu$m the predicted corrections are not far from the TR-2 ($\sim$12\% spot filling factor) case. The chromatic dependence, however, is significantly different. It is therefore possible that the transmission spectrum of \cite{Bruno_2019} is somewhat affected by the different chromatic slope, although the average correction value should not be strongly biased. Note, however, that if we were to use the spot parameters adopted by \cite{Bruno_2019} ($\Delta T_{\rm sp}$=2200~K, compared with 575~K in our work) at NIR wavelengths, very strong deviations would occur, with spot corrections overestimated by factors of 2 to 5.

The differences found highlight the importance of extracting the chromatic signature from multiband photometric datasets so that the degeneracy between spot temperature and filling factor can be broken, and to reproduce realistic spot distributions. The determination of physical properties of active regions is a key point for the study of their chromatic signature on transit observations. Besides, using independent data, instead of the same transmission spectroscopy used for exoplanet characterisation, prevents misinterpreting atmospheric features.

\begin{figure}
\begin{center}
\includegraphics[width=9cm]{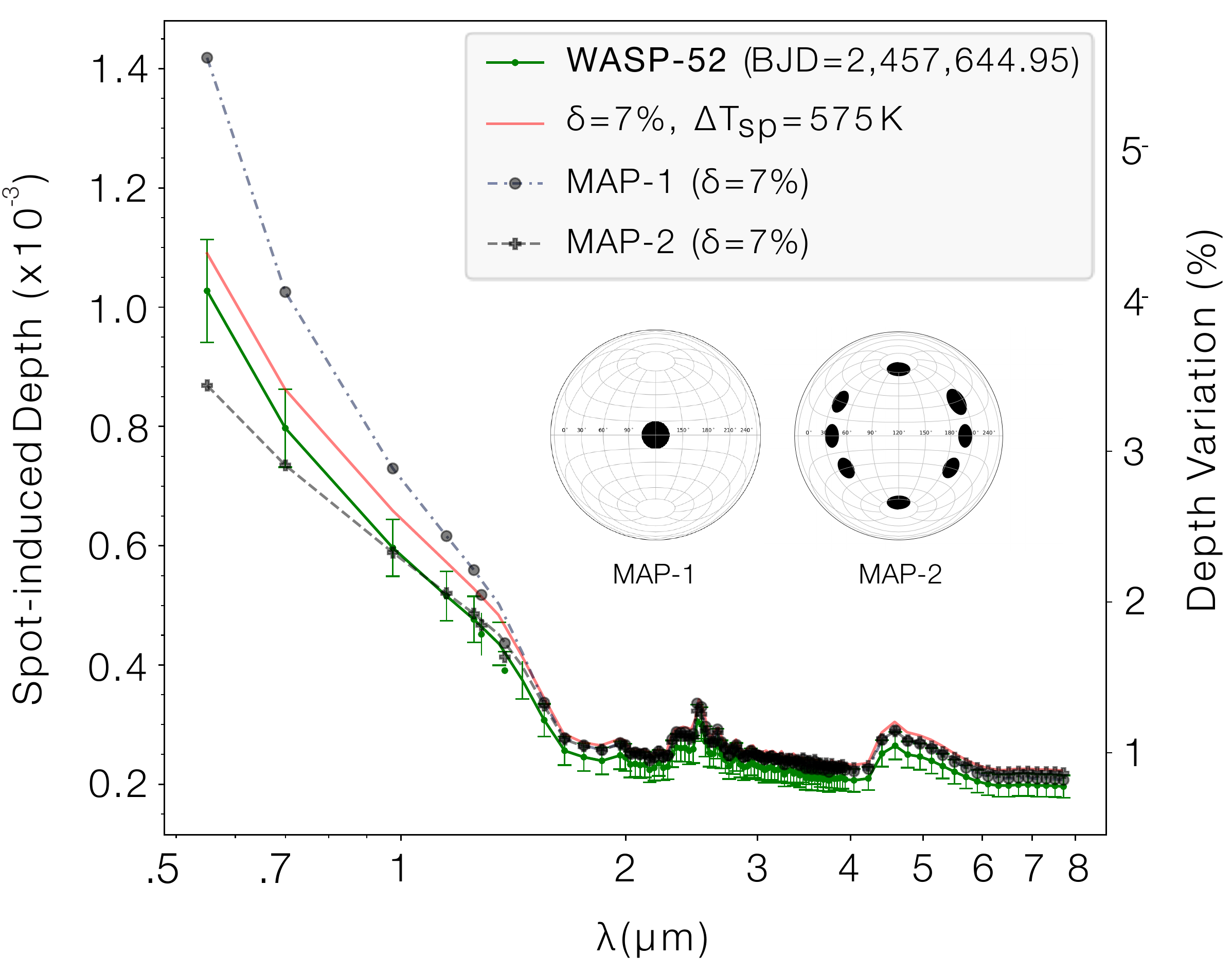}
\end{center}
\caption{Chromatic effects due to spot distributions with identical filling factors. MAP-1 and MAP-2 are two ad-hoc generated spot maps to show the differences in their depth profiles and the comparison with that provided with Eq.\,(\ref{eq:rackham}), shown in red. The green curve is a fitted map for WASP-52 at epoch BJD=$2,457,644.95$ with the parametric distribution error bands.}
\label{fig:LD}
\end{figure}

We caution that the procedure employed in the literature above corrects the chromatic effects of active regions in transits by estimating the spot filling factor and temperature. However, the actual spot distribution on the stellar surface is still relevant because of limb darkening effects. Figure \ref{fig:LD} shows a comparison between two ad-hoc spot maps (MAP-1 and MAP-2), a fitted map of WASP-52 at epoch BJD=2,457,645 and the theoretical depth variation with homogeneous distribution (Eq.\,\ref{eq:rackham}). All these models have been calculated with the same projected spot coverage of 5\%. MAP-1 has a single spot at the centre of the disc while MAP-2 has spot group close to the limb. The differences are quite significant, especially in the visible. This simple exercise illustrates the need to consider the full stellar surface at the time of transit to properly correct for spot effects. 

\section{Conclusions} \index{S7} \label{S7}

In this study we present the \textsf{StarSim} code, which is able to model the surface of active stars through the inversion of photometric time-series. With the resulting model, we can investigate the influence of the chromatic effect produced by activity features for transmission spectroscopy. As a test case, we use simultaneous $BVRI$-filter photometry of the young active planet-host star WASP-52 covering $\sim$600 days of 2016 and 2017 from the STELLA and TJO telescopes. The data are used to constrain an activity model with a heterogeneous surface composed of dark spots with a temperature difference of $575\pm150$ K with respect the surrounding photosphere. Our modelling rejects the hypothesis of the presence of bright facular regions, as expected for young FGK stars. We determine the rotation period of the star to be $17.26^{+0.51}_{-0.39}$ days. The fit also yields a probabilistic map of active regions, showing two or three prominent spot complexes at stable stellar longitudes.

It is important to emphasise that the ultimate goal of our effort is to correct the transit depth as a function of wavelength, and not necessarily to obtain precise reconstructions of the stellar surface features. That is, although there might be degeneracies in the number of spots and their exact distribution, it is the overall photometric effect at the time of transit what really matters. To study the influence of the derived activity model on the observables of transmission spectroscopy, we produced simulated transits of WASP-52~b by avoiding spot-crossing events. The chromatic effect of the spot map results in uncertainties of $\sim$10\% in transit depth (SP($\lambda$)/$\mathcal{D}_0$) and of up to $\sim$5\% in the planetary radius ($\Delta r_{\rm SP}/r$), at visible wavelengths. After correcting for spot effect using the \textsf{StarSim} model, we are able to reduce residual depth uncertainties down to $\sim$10$^{-4}$ at 550\,nm (Ariel/VIS-Phot) and $\sim$$3\times10^{-5}$ at 6\,$\mu$m (Ariel/AIRS-Ch1). 

The remaining uncertainty of the correction factor, computed as the 1-$\sigma$ interval of the depth variation of the simulated transits for the fitted spot maps, encloses several effects, most importantly: 1) the photometric precision and phase coverage of the monitoring along time, 2) the incompleteness of the model implemented and, 3) the uncertainties in retrieving optimal surface maps and stellar parameters in the context of the photometric inverse problem.

The main advantage of the approach we present to compute the chromatic effect on the depth of exoplanet transits is that it is based on an independent and consistent deterministic method allowing to accurately determine the stellar parameters, the filling factor and the distribution of spots. Thus, the effect of spots and their different positions on the stellar disc are also taken into account. We have only considered here the effect of unocculted spots, which increase the transit depth. On the other hand, occulted spots produces the opposite effect. In our previous work on GJ~1214, we described the difficulty to achieve a unique solution for the planetary optical spectral slope when the transit observation is affected by a mixture of non-occulted and occulted spots, even if multi-colour monitoring of the star is available \citep{Mallonn_2018}. However, very-high precision transit photometry could help to detect the effects of occulted spots and to derive their properties.

With the expected launch of JWST in 2021 and Ariel mission in 2028, the ability to observe transmission spectra of transiting planets will increase both in the number of stars and precision. In this context, our results show that contemporaneous ground-based photometric monitoring will be crucial to adequately model spot-driven stellar activity variations and correct out their chromatic effects on the transit depth measurements. In essence, our method exploits the chromaticity of spot-induced stellar time variability. Activity correction may be essential to attain the required accuracy in transmission spectroscopy so that meaningful constraints on the planetary atmospheres can be defined. We note that the correction is necessary irrespective of the precision of the space-based photometry since it deals with a systematic effect of astrophysical origin. We expect that higher activity attenuation factors could be achieved by optimizing the strategy of the photometric observations, that is, making sure that multi-colour measurements are obtained before and after the transits to better constrain the model. Photometric observations from space would even boost the accuracy in light curve modelling, and hence, the correction capabilities. 

\begin{acknowledgements}
We are grateful to the referee for their valuable and insightful comments and suggestions that helped to improve the paper significantly. We acknowledge support from the Spanish Ministry of Science and Innovation and the European Regional Development Fund through grant PGC2018-098153-B-C33, as well as the support of the Generalitat de Catalunya/CERCA programme.
\end{acknowledgements}

\begin{appendix}

\end{appendix}

\bibliographystyle{aa}
\bibliography{biblio}

\end{document}